\documentclass[12pt]{article}
\usepackage{multicol,amscd}

\usepackage{hyperref}

\usepackage{amsthm,amsmath,latexsym,amssymb,amsfonts}

\newcounter{Summa}
\newcounter{Csx}
\newcounter{Csy}

\newcommand{\YD}{Young diagram\quad}
 \newcommand{\nn}{\nonumber}
\newcommand{\Length}[1]{#10}

\newcommand{\CellSize}{10}

\newcommand\ls{\!\!\!\!\!\!\!}

\newcommand{\Inc}[2]{\setcounter{#1}{#2} \addtocounter{#1}{1} }

\newcommand{\YoungBlock}[2] {
    \begin{picture}(\Length{#1},\Length{#2})(0,0)
    \Inc{Csx}{#1} \Inc{Csy}{#2}
    \multiput(0,0)(\CellSize,0){\value{Csx}}{\line(0,1){\Length{#2}}}
    \multiput(0,0)(0,\CellSize){\value{Csy}}{\line(1,0){\Length{#1}}}
    \end{picture}
  }

\newcommand{\YoungGeneralized }
{    \begin{picture}(140,100)(0,0)
    \put(0,80) { \YoungBlock{11}{1} $L_1$}
    \put(0,70) { \YoungBlock{9}{1} $L_2$}
    \put(0,60){ \YoungBlock{8}{1} $L_3$}
    \put(4,30){ \line(0,1){30}}
    \put(10,47){$\vdots$}
    \put(10,33){$\vdots$}
    \put(0,20) { \YoungBlock{4}{1} $L_{H-2}$}
    \put(0,10) { \YoungBlock{4}{1} $L_{H-1}$}
    \put(0,0){ \YoungBlock{1}{1} }
    \put(05,-13){$H_1$ }
    \put(18,-13){$H_2\,\,\,\dots$ }
    \put(28,0){$L_{H_1}$ }
    \end{picture}
}

\newcommand{\YoungGeneralDifBl}{
{    \begin{picture}(140,150)(0,80)
\put(7,210){\linethickness{0.4mm}\line(1,0){170}}%
\put(7,211){$\overbrace{\rule{170pt}{0pt}}$} \put(80,222){$L_1$}
\put(-15,195){$F_1$}

\put(0,197){$\left\{\rule{0pt}{11pt}\right.$}

    \put(0,190) { \YoungBlock{17}{2} }
\put(7,190){\linethickness{0.4mm}\line(1,0){170}}%
\put(7,141){$\overbrace{\rule{110pt}{0pt}}$} \put(60,152){$L_2$}

\put(-15,160){$F_2$} \put(-2,162){$\left\{\rule{0pt}{27pt}\right.$}
    \put(0,140) { \YoungBlock{11}{5} }
\put(7,140){\linethickness{0.4mm}\line(1,0){110}}%
\put(7,109){$\underbrace{\rule{80pt}{0pt}}$} \put(48,90){${  L_3}$}
\put(-15,120){$F_3$} \put(0,122){$\left\{\rule{0pt}{18pt}\right.$}
    \put(0,110){ \YoungBlock{8}{3} }
\put(7,110){\linethickness{0.4mm}\line(1,0){80}}%
\put(7,110){\linethickness{0.4mm}\line(0,01){100}}
\put(87,110){\linethickness{0.4mm}\line(0,01){30}}%
\put(117,140){\linethickness{0.4mm}\line(0,01){50}}%
\put(177,190){\linethickness{0.4mm}\line(0,01){20}}%
    \end{picture}
} }

\newcommand{\YoungGeneralDifBll}{
{    \begin{picture}(200,150)(0,80)

\put(-15,195){$F_1$}

\put(0,197){$\left\{\rule{0pt}{11pt}\right.$}

    \put(0,190) { \YoungBlock{17}{2} }

\put(-15,160){$F_2$} \put(-2,162){$\left\{\rule{0pt}{27pt}\right.$}
    \put(0,140) { \YoungBlock{11}{5} }
\put(-15,120){$F_3$} \put(0,122){$\left\{\rule{0pt}{18pt}\right.$}
    \put(0,110){ \YoungBlock{8}{3} }
\put(182,200){$L_1$} \put(120,178){$L_2$} \put(95,120){${  L_3}$}

    \end{picture}
} }

\newcommand{\YGBR}{
{    \begin{picture}(200,150)(0,80) \put(183,200){$L_1$}
\put(159,187){$\tilde{L}_1$}

\put(2,201.5){$\left\{\rule{0pt}{4pt}\right.$} \put(0,200) {
\YoungBlock{17}{1} } \put(0,190) { \YoungBlock{15}{1} }

\put(-35,200){$F_1\!-\!1$}

\put(-35,160){$F_2\!-\!1$}
\put(-0,166.5){$\left\{\rule{0pt}{22.5pt}\right.$} \put(0,150) {
\YoungBlock{11}{4} } \put(0,140) { \YoungBlock{9}{1} }

\put(120,178){$L_2$} \put(101,137){$\tilde{L}_2$}

  \put(95,120){${  L_3}$}
  \put(31,107){${  \tilde{L}_3}$}
\put(-35,120){$F_3\!-\!1$}
\put(2,126){$\left\{\rule{0pt}{11pt}\right.$}
    \put(0,120){ \YoungBlock{8}{3} }
    \put(0,110){ \YoungBlock{2}{1} }
    \end{picture}
} }

\newcommand{\pls}
{\begin{picture}(14,10)(7,1)
 {\linethickness{1pt}
\put(3.5,0) { \line(1,0){11}}
\put(3.5,10) { \line(1,0){11}}
\put(4 ,0) { \line(0,1){10}}
\put(14 ,0) { \line(0,1){10}}
 }
 \put(10.5,5) {\linethickness{0.4mm}\line(1,0){5}}
  \put(13,2.5) {\linethickness{0.4mm}\line(0,1){5}}
 \end{picture}}

 \newcommand{\mns}
 {\begin{picture}(14,10)(7,1)
{\linethickness{1pt}
\put(3.5,0) { \line(1,0){11}}
\put(3.5,10) { \line(1,0){11}}
\put(4 ,0) { \line(0,1){10}}
\put(14 ,0) { \line(0,1){10}}
 }
\put(10.4,5) {\linethickness{0.4mm}\line(1,0){5}}
  \end{picture}}

\newcommand{\krs}
 {\begin{picture}(14,10)(7,1)
{\linethickness{1pt}
\put(3.8,0) { \line(1,1){10}}
\put(3.8,10) { \line(1,-1){10}}
 }
  \end{picture}}

\newcommand{\klt}
 {\begin{picture}(14,10)(7,1)
{\linethickness{1.2pt}
\put(3.5,0) { \line(1,0){11}}
\put(3.5,10) { \line(1,0){11}}
\put(4 ,0) { \line(0,1){10}}
\put(14 ,0) { \line(0,1){10}}
 }
\end{picture}}

\newcommand{\mklt}
 {\begin{picture}(14,10)(1,5)
{ \linethickness{0.4mm}
\put(3.5,0) { \line(1,0){7}}
\put(3.5,7) { \line(1,0){7}}
\put(4 ,0) { \line(0,1){7}}
\put(11 ,0) { \line(0,1){7}}
 }
\end{picture}}

\newcommand{\mlv}
 {\begin{picture}(14,10)(7,1)
{\linethickness{.4mm}
\put(3.5,-0) { \rule{10pt}{10.2pt}}
 }
\end{picture}}

\newcommand{\col} 
{    \begin{picture}(30,25)(-10, 5)
{ \linethickness{0.4mm}
\put(-5,13){${}_p$}

\put(0,9){$\left\{\rule{0pt}{20pt}\right.$}
\put(9.6,7.5){$\vdots$}
\put(10.5,-5){ \line(0,1){35}}
\put(3.5,-5){ \line(0,1){35}}

    \put(0.5,0) {\mklt}
       \put(0.5,28) {\mklt}
 }   \end{picture}
}

\makeatletter \@addtoreset{equation}{section} \makeatother

\def\be{\begin{equation}}
\def\ee{\end{equation}}
\newcommand{\V}{{\cal V}}

\def\bee{\begin{eqnarray}}
\def\eee{\end{eqnarray}}

\def\bal{\begin{align}}
\def\eal{\end{align}}

\newcommand{\W}{\mathcal{W}}

\def\ba{\begin{array}}
\def\ea{\end{array}}
\newcommand{\ie}{{\it i.e.,} }
\def\nn{\nonumber}
\newcommand{\f}{\frac}
\def\D{\mathcal{D}}

\newcommand{\half}{\frac{1}{2}}

\def\p{\partial}

\newcommand{\q}{\,,\qquad}

\newcommand{\+}
{\begin{picture}(14,10)(7,1)
 {\linethickness{1pt}
\put(3.5,0) { \line(1,0){11}}
\put(3.5,10) { \line(1,0){11}}
\put(4 ,0) { \line(0,1){10}}
\put(14 ,0) { \line(0,1){10}}
 }
 \put(10.5,5) {\linethickness{0.4mm}\line(1,0){5}}
  \put(13,2.5) {\linethickness{0.4mm}\line(0,1){5}}
 \end{picture}}

 \newcommand{\mi}
 {\begin{picture}(14,10)(7,1)
{\linethickness{1pt}
\put(3.5,0) { \line(1,0){11}}
\put(3.5,10) { \line(1,0){11}}
\put(4 ,0) { \line(0,1){10}}
\put(14 ,0) { \line(0,1){10}}
 }

 \put(10.4,5) {\linethickness{0.4mm}\line(1,0){5}}
  \end{picture}}

  \def\cC{\mathcal{C}}
  \def\cF{\mathcal{F}}
 \def\cH{\mathcal{H}} 
  \def\cL{\mathcal{L}}
  
  \def\cR{\mathcal{R}}

\newcommand{\ga}{\alpha}
\newcommand{\gb}{\beta}

\newcommand{\go}{\omega}    
\newcommand{\gs}{\sigma}
\newcommand{\bL}{\mathbf{L}}

\newcommand{\um}{{\underline m}}
\newcommand{\un}{{\underline n}}

\newcommand{\uk}{{\underline k}}
\newcommand{\ul}{{\underline l}}
\newcommand{\B}{{ B}}
\newcommand{\CV}{{ V^\natural}}

\begin{document}

\begin{flushright}
{\small FIAN/TD/15-09}
\end{flushright}

\vspace{1mm}

\begin{center}

{\bf \Large Bosonic conformal higher--spin  fields\\
 of any symmetry}

\vspace{1.2cm}

M.A. Vasiliev

\vspace{5mm}

I.E. Tamm Department of Theoretical Physics,\\ P.N. Lebedev Physical Institute, \\
Leninsky prospect 53, 119991 Moscow, Russia

\vspace{18mm}

\hspace{6.5cm}\textit{To the memory of Efim Samoilovich Fradkin }

\vspace{18mm}

\vspace{8mm}
\begin{abstract}
Free Lagrangians  are found both for gauge and  non-gauge
bosonic conformal fields of any symmetry type and in any space-time
dimension. Conformal gauge fields of various types,
their gauge transformations and gauge invariant field strengths
(generalized Weyl tensors), which are derived by the $\sigma_-$
cohomology
technics in the frame-like formulation, are shown to
correspond to supersymmetric vacua of certain supersymmetric matrix
mechanics. The correspondence between
conformal and $AdS_d$ higher-spin models, that turn out to have
identical generalized Weyl tensors, is discussed.

\end{abstract}

\end{center}

\newpage

{\footnotesize \tableofcontents}

\newpage

\section{Introduction}

Symmetric conformal higher--spin (HS) gauge fields were
originally studied by Fradkin and Tseytlin in \cite{FT} where
gauge invariant actions for these
fields  were found in four dimensions. These authors used
so-called metric-like formalism developed originally for
Poincar\`{e} invariant unitary relativistic systems including
symmetric massive HS fields \cite{singh&hagen}, symmetric
massless HS fields \cite{Fm0} and massless fields of arbitrary
symmetry type \cite{Labastida}. (For related work see also
\cite{Francia:2002aa,Bekaert:2002dt,Zinoviev:2002ye,de
Medeiros:2003dc,Buchbinder:2007ix,
Francia:2007ee,Buchbinder:2008ss,Reshetnyak:2008sf,Campoleoni:2008jq}.)
 The results of \cite{FT} were
extended to any dimension by Segal \cite{segal} and to mixed
symmetry gauge fields described by rectangular Young tableaux of height
$n$ in even dimension $d=2(n+1)$ by Marnelius \cite{Marnelius:2009uw}.

Specific examples of conformal HS fields have been extensively
studied in the literature (see in particular
\cite{DW,AM,Metsaev:2007rw,Metsaev:2007fq,mets}). After the works
\cite{Siegel,Mets} it is known that, beyond four dimensions, the
class of conformal fields that corresponds to unitary theories is
very restricted. Namely, apart from massless scalar and spinor in
any dimension, only mixed symmetry fields that have field
strengths described by rectangular Young diagrams of height $d/2$
in even space--time dimensions, studied in
\cite{town,hull,BBAST,bast}, have unitary spaces of single particle
states. However, the  formulation of these systems in terms of
gauge potentials breaks down conformal invariance as happens, for
example, for the $4d$ massless spin two field described in terms
of metric and, more generally,  for all  gauge fields except for
the generalized ``spin one" fields  described by rank $d/2-1$
differential form potentials in even dimension, which include
usual $4d$ spin one\footnote{ Note that in \cite{33} it was shown
that this effect can be avoided in the $AdS$ geometry, that is the
conformal invariant formulation of \cite{33} for general $4d$
gauge fields becomes singular in the flat limit.}. Beyond this
class,  conformal models, that allow gauge invariant (and hence
Lagrangian) formulation in  flat space, contain higher derivatives
leading to ghosts. An example of such a model is provided by the
$4d$ Weyl gravity with the Lagrangian (see, e.g., \cite{FT}) \be L
=C_{nm,kl}C^{nm,kl}\,, \ee where $C_{nm,kl}$ is the Weyl tensor.
In this case, the field equations contain four derivatives.

In this paper we describe bosonic conformal mixed
symmetry fields of general type in Minkowski space, using geometric
methods of the frame-like formulation and unfolded dynamics
approach. The frame-like formulation was originally proposed for the
description of symmetric massless fields of any spin in flat
\cite{V80} and $AdS$ spaces \cite{V1,LV} and then extended to $4d$
conformal symmetric massless fields in \cite{FL}, to mixed
symmetry fields in $AdS_4$ \cite{ASV1,ASV2} and Minkowski
\cite{skmin} spaces, as well as to string inspired reducible sets
of fields \cite{SORV}, partially massless fields \cite{SV} and even to
massive fields \cite{mf}.
Unfolded formulation\footnote{This terminology was introduced in
\cite{un}, while the approach was originally introduced and
applied to the analysis of HS theory in
 \cite{VA,Ann}.} is a
specific reformulation of partial differential equations in a
coordinate--independent first-order form by virtue of introducing an
appropriate (may be infinite) set of auxiliary fields.

In fact, the full list of conformal invariant equations has been
already elaborated in \cite{STV} using the unfolded dynamics
approach. However, being formulated in terms of zero--forms, the
approach of \cite{STV} makes gauge symmetries not manifest. In this
paper we apply unfolded dynamics to the frame-like formulation
with manifest gauge symmetries of differential form  gauge
fields, which reformulation is  important for the further
analysis of interactions. In addition, we construct free conformal
invariant actions that describe both gauge and non-gauge mixed
symmetry conformal fields of general type.

As known from supergravity \cite{sugr,FT} and world-like particle
models \cite{Marn}, (super)conformal models provide a useful tool
for the study of quantum-mechanically well-defined
unitary models in terms of spontaneously broken conformal symmetry. A
version of this approach is known as two-time physics
\cite{B}. In all cases, the idea is that it is useful to describe
non-conformal models as spontaneously broken conformal ones. We
expect  that analogous phenomenon is true for HS theories
and the results of this paper will be useful for the study of the
unitary HS models.

In the unfolded dynamics approach, dynamical content of a model is
characterized by the so-called $\gs_-$
cohomology \cite{sigma} (for recent discussions and reviews see
\cite{solv,act,BIS}) which, in particular, classifies all
gauge invariant tensors that can be constructed from a given gauge
field. From
this perspective, the important difference between  conformal models and
those in $AdS_d$ is that the analysis of the former is based on the
study of $\gs_-$ complex that has clear group--theoretical meaning with
respect to the conformal algebra $o(d,2)$, that, in turn, greatly
simplifies the study of its cohomology. In the unitary $AdS_d$ case,
the $\gs_-$ complex is more involved, having no direct
group--theoretical interpretation in terms of the $AdS_d$ algebra
$o(d-1,2)$. The idea is that the interpretation of
$o(d-1,2)$ as a subalgebra of
the conformal algebra $o(d,2)$ provides  natural relation between
the $AdS$ and  conformal $\gs_-$ cohomology. In fact, one of
the motivations for the study of this paper is  to interpret
$AdS_d$ cohomology, obtained recently in \cite{skads1}, in terms
of the conformal algebra cohomology. This not only gives
an efficient tool for the study of formal aspects of the
HS gauge theory but, hopefully, will lead to its
``compensator version" with spontaneously broken conformal symmetry.
Eventually, accomplishment of this program  may have important
applications to the study of nonlinear HS theories with mixed symmetry fields.

A useful observation that greatly simplifies the computation is that
 the $\gs_-$ cohomology in conformal
gauge theories consists of supersymmetric vacua of some
 supersymmetric matrix mechanics. We explain
this method in some detail because it can be applied to other
interesting physical models associated to $sl_n$ and $sp(2M)$
symmetric models. The $sl_n$ case is expected to be related to
off-shell higher--spin theories of the type considered in the
pioneering paper \cite{DWF}. The case of $sp(2M)$ corresponds to
HS gauge theories in the generalized space-time with
symmetric matrix coordinates \cite{F,BLS,BHS,Mar}.

The idea to describe a conformal
invariant system in the manifestly $o(d,2)$ covariant way  was
originally proposed by Dirac in his celebrated paper \cite{dcon}
that underlies most of the modern approaches to conformal field
theories beyond two dimension (see \cite{Marn,con,B,AM} and references
therein). In mathematical literature, a similar approach is called
tractor theory \cite{East,gov} (for applications and more references see
\cite{GSW}.)
A new element of the unfolded dynamics approach is that it operates
with differential forms valued in an
irreducible $o(d,2)$--module $V$. In this paper we consider the
important  subclass of finite dimensional $o(d,2)$--modules, \ie
spaces of irreducible tensors of $o(d,2)$.
Comparing obtained results with those of \cite{STV} we shall see that this class
contains all possible finite-component conformal gauge fields.
The extension to differential forms gives several
benefits. One is that space-time enters the
construction rather implicitly. The only characteristics, that
 matters in practice, is its dimension. In other words, it is not
important how exactly space-time is realized: as a slice of the
projective cone or simply as Minkowski space. In all cases, conformal
invariance follows from the frame-like formulation in terms of
$o(d,2)$--modules.

The structure of the paper is as follows.

Section \ref{prel} recollects  some
well-known facts. In Subsection \ref{Conformal algebra}, we introduce
conformal algebra notation. In Subsection \ref{civ}, we recall
how conformal invariant backgrounds and global conformal
symmetries result from the zero-curvature equations for $o(d,2)$.
Main elements of the $\sigma_-$ cohomology analysis of
unfolded  partial differential equations are summarized in Subsection
\ref{sigma-}. In Subsection \ref{Fock}, we recall how  tensorial
spaces can be described as subspaces of appropriate Fock spaces.

Section \ref{Sigma} contains the analysis of $\sigma_-$ cohomology
for conformal gauge theories as well as its dynamical interpretation. The
operator $\sigma_-$ for the finite dimensional conformal modules is
introduced in Subsection \ref{sigcon}. The homotopy operator is
introduced in Subsection \ref{homop} while its
interpretation in terms of supersymmetric matrix mechanics is
given in Subsection \ref{sup}. The $\sigma_-$ cohomology is computed in
Subsection  \ref{comp} and its dynamical interpretation is presented in
Subsection \ref{dint}. Subsection \ref{dcs} contains illustrative examples.
General structure of the unfolded field equations for conformal gauge fields
is discussed in Subsection \ref{stue}.

Conformal invariant actions for gauge
and non-gauge fields are worked out in Section \ref{act}
which is rather independent. The shortcut is through
Subsections \ref{Conformal algebra},
\ref{civ} and \ref{dint}-\ref{stue}.

Section \ref{weylmod} focuses on the interpretation
of the obtained results in terms of infinite dimensional modules of
conformal algebra in the context of the results of \cite{STV} and
possible application of the unfolded dynamics at the action level.

Conclusions and perspectives are discussed in Section \ref{conclusion}
with the emphasize on the relation of the obtained results with
the unitary HS models.

Appendix presents some properties of the  conformal invariance condition.

\section{Preliminaries}
\label{prel}

\subsection{Conformal algebra}
\label{Conformal algebra}

Generators $T^{AB}$ of $o(d,2)$ satisfy the commutation relations
\be \label{d2c} [T^{AB}\,,T^{CD}]=\eta^{BC} T^{AD} - \eta^{AC}
T^{BD} - \eta^{BD} T^{AC} + \eta^{AD} T^{BC}\,,
\ee
where $\eta^{AB}=\eta^{BA}$ is a
nondegenerate $o(d,2)$ invariant symmetric form.
Indices
$A,B,\ldots $ take $d+2$ values and are raised
and lowered by $\eta^{AB}$ and its inverse
$\eta_{AB}$. We arrange
$A=(a,-,+)$ with $a,b,\ldots =  0,\ldots d-1$
so that
\be \eta^{+-}=1\q \eta^{++}=\eta^{--}=0\q
\eta^{\pm a}=0\,, \ee
and $\eta^{ab}$ is a nondegenerate  invariant form of the Lorentz algebra
$o(d-1,1)$ in $d$ dimensions. In Lorentz notation we assign
\be \label{lorid}
T^{-a} = P^a\q T^{+a} = K^a\q T^{+-} = D\q T^{ab}=L^{ab}\,, \ee
where $P^a$, $K^a$, $D$ and $L^{ab}$ are the generators of
translations, special conformal transformations, dilatations and
Lorentz transformations, respectively. Note that $D$ induces the
following grading of the algebra
\be
[D\,,P^a]= -P^a\q [D\,,K^a]=
K^a\q [D\,, T^{ab} ]=0.
\ee

Conformal gravity can be described in terms of
$o(d,2)$ one-form connection
\be W (x)= \half W_{\,AB}(x)T^{AB}\q W_{\,AB}(x) = dx^\un
W_{\un\,AB}(x) \ee
and two-form curvature
\be \ls\,\,\,\, R (x)= \half R_{\,AB}(x)T^{AB}\q\!\! R_{AB}(x)= dW_{\,AB}(x)
+W_{\,AC}(x)\wedge W_{}^C{}_B(x)\q \!\!\!\!\!d=dx^\un \f{\p}{\p x^\un}\,.
\ee
(Underlined indices
$\um, \un =0,\ldots d-1$ are associated to vector fields and
differential forms on the base manifold.)
 In terms of Lorentz irreducible components we have
 \be
\label{congau} W(x)= h^a(x) P_a +\half \go^{ab}(x) L_{ab} + f_a
(x)K^a + b(x) D\,,
\ee
\be R(x)= R^a (x)P_a +\half R^{ab}(x) L_{ab} + r_a (x)K^a + r(x)
D\,,
\ee
where \be\label{Rh} R^a = dh^a +\go^a{}_b\wedge h^b
-b\wedge h^a\,, \ee \be\label{Rgo} R^{ab} = d\go^{ab}
+\go^a{}_c\wedge \go^{cb} -h^a\wedge f^b+ h^b\wedge f^a\,, \ee
\be\label{Rb} r= db +h^a \wedge f_a\,, \ee \be\label{Rf} r^a = df^a
+\go^a{}_b\wedge f^b + b\wedge f^a\,. \ee

Here $\go^{ab}(x)$ is Lorentz connection, $f_a(x)$ and $b(x)$ are gauge
fields for special conformal transformations and dilatation,
respectively, and $h^a=dx^\um h_\um{}^a$  is the vielbein one-form that
is required to be nondegenerate
\be
\label{nondeg}
det  | h_\um{}^a |\neq 0\,.
\ee

The conformal gauge transformations are
\be\label{dh} \delta h^a =
\D^L\epsilon^a -\epsilon^a{}_b h^b +\epsilon  h^a -\epsilon^a b\,,
\ee \be \label{dgo} \delta\go^{ab} = \D^L\epsilon^{ab} -h^a
\tilde{\epsilon}^b +\epsilon^a f^b +h^b \tilde{\epsilon}^a
-\epsilon^b f^a\,, \ee
\be \label{db} \delta b = d \epsilon + h^a
 \tilde{\epsilon}_a - \epsilon^a f_a\,,
\ee
\be\label{df}
\delta f^a = \D^L \tilde{\epsilon}^a -\epsilon^a{}_b f^b -\epsilon
f^a +\tilde{\epsilon}^a b \,,
\ee where $\epsilon^a(x)$,
$\epsilon^{ab}(x)$, $\tilde{\epsilon}_a(x)$ and $\epsilon(x)$ are
gauge parameters of translations, Lorentz transformations, special
conformal transformations and dilatations, respectively. $\D^L$ is
the Lorentz covariant derivative \be \D^L \psi^a = d\psi^a
+\go^a{}_b\wedge \psi^b\,. \ee

The $AdS_{d}$ algebra $o(d-1,2)$ can be realized as the subalgebra
of $o(d,2)$ spanned by the generators $L_{ab}$ and $ {\cal P}_a =
P_a +\lambda^2 K_a $, where $\lambda^2>0$ ($\lambda^2
<0$ corresponds to the de Sitter case of $o(d,1)\subset o(d,2)$).
The limit $\lambda \to 0$ gives the Poincare subalgebra
$iso(d-1,1)\subset o(d,2)$.

\subsection{Unfolded dynamics}
\label{civ}

In the unfolded dynamics approach \cite{Ann}
(for more detail see e.g. \cite{solv,33}), $g$--invariant dynamical
systems are described in terms of differential forms $W^\Omega$
 valued in one or another $g$-module (index $\Omega$). In
the case of most interest in this paper, $g=o(d,2)$.

A $g$--invariant background is described by a flat connection of
$g$, \ie a one-form $W_0$ that satisfies the zero--curvature
equation
\be \label{r0} dW_0 +\half [W_0\,,W_0]=0\,,
\ee
where
$[\,,]$ is a Lie bracket in $g$. Usually, $g$ contains one or
another space-time symmetry as a subalgebra. The part of $W_0$
associated to the Poincar\`{e} or $AdS$ translations is
identified with the frame one--form.

In the case of $g=o(d,2)$, different vacuum solutions
$W_0(x)$ may have different interpretations.
If  nonzero components of $W_0$ are in the Poincar\`{e} subalgebra
$iso(d-1,1)\subset o(d,2)$ then it describes (locally) Minkowski
space-time. For example, this is true in the case of
Cartesian coordinates in Minkowski space where the only
nonzero component of $W_0$ is
\be \label{mink}
h^a = dx^a\,.
\ee

If  nonzero components of $W_0$ belong to the $AdS_d$ subalgebra
$o(d-1,2)\subset o(d,2)$, then it describes (locally) $AdS_d$
space-time.
A flat connection of $o(d-1,2)$, namely $e^a$ and
$\go^{ab}$, gives a flat connection of $o(d,2)$ with $ h^a =
 e^a$, $ f_a = \lambda^2 e_a$, $ b=0$ and the same Lorentz
connection $\go^{ab}$, \ie this Ansatz solves (\ref{r0}) for the
conformal algebra provided that $e^a$ and $\go^{ab}$ solve the
zero curvature
equations for $o(d-1,2)$. de Sitter case is described analogously
with $\lambda^2 <0 $.)
 Note that, as a by-product, this gives a
coordinate independent proof of conformal flatness of $(A)dS_d$.

Coming back to generic $g$-symmetric case, let $\D_0$ be a
$g$--covariant derivative built from $W_0$. Then (\ref{r0}) implies
\be \label{d02} \D_0^2=0\,. \ee
This allows us to introduce a
linearized curvature
\be \label{R1} R_1^\Omega = \D_0
W_1^\Omega\,
\ee for any $p$-form $W_1^\Omega$ valued in some
$g$-module $V$. Because of (\ref{d02}), $R_1^\Omega$
satisfies the Bianchi identities
\be \label{bi} \D_0 R_1^\Omega (x)=0
\ee
 and is gauge
invariant under the gauge transformation
\be \label{gtr} \delta
W^\Omega_1(x) = \D_0\epsilon^\Omega_{p-1} (x)\,,
\ee
where
$\epsilon^\Omega_{p-1}(x)$ is
an arbitrary $V$--valued ($p-1$)--form. Also note that there is a
chain of gauge symmetries for gauge symmetries
\be \label{hgtr}
\delta\epsilon^\Omega_i(x) =\D_0
\epsilon^\Omega_{i-1}(x)\q 0\leq i\leq p-1\,,
\ee
that leave invariant the gauge transformations (\ref{gtr}) and
 the transformations (\ref{hgtr}) themselves.

We can let $W_1$  describe a set of differential
forms of different degrees extending the definition (\ref{R1}) to
\be
\label{Rco} R^\Omega_1 = d W_1^\Omega + F^\Omega (W_0, W_1)\,,
\ee
 where
$F^\Omega (W_0, W_1)$ is built from wedge products of the differential forms
$W_0$ and $W_1$ and is linear in $W_1$, but not necessarily in
$W_0$. The equation
\be \label{freq} R^\Omega_1=0 \ee
is required to be formally
consistent  with (\ref{r0}), \ie to respect
 $d^2=0$ for any $W_0$ that satisfies (\ref{r0}). The system (\ref{r0})
 and (\ref{freq}) expresses the exterior
differential of any of the fields $W_1^\Omega(x) $ and $W^A_0$ via wedge
products of the same set of fields. Such differential equations are called
{\it unfolded}.

General unfolded systems are associated to the curvatures
\be
R^\ga=d\W^\ga+F^\ga (\W)
\ee
where $\W^\ga(x)$ is some set of differential forms  and
$F^\ga (\W)$ obeys the generalized Jacobi condition
\be
\label{conci}
F^\ga (\W) \f{\p F^\gb(\W)}{\p \W^\ga}=0\,,
\ee
that expresses the compatibility of the unfolded equation
\be
\label{genunf}
R^\ga (\W(x))=0
\ee
with $d^2=0$.
The important subclass of unfolded systems is constituted by
the so-called {\it universal} unfolded systems \cite{solv} where (\ref{conci})
still takes place with the odd and even differential form variables
treated as odd and even supercoordinates of some superspace. In other
words, the unfolded system is universal if (\ref{conci}) is
insensitive to the dimension of space-time $d$ (\ie to the property that
any ($d+1$)--form is zero; for more detail see \cite{33}). The
unfolded systems that appear
in HS models, and, in particular, the systems studied
in this paper are universal. In this case the derivative
$\frac{\p F^\Omega(\W)}{\p \W^\ga}$ is well-defined,
allowing to
define the transformation law \cite{Ann}
\be
\label{ggtr}
\delta \W^\ga (x) = d \epsilon^\ga (x)-
\epsilon^\gb (x) \f{\p F^\ga(\W(x))}{\p \W^\gb(x)}\,,
\ee
where the derivative is left and
$\epsilon^\ga (x)$ is an arbitrary ($p_\ga-1$)--form if $\W^\ga(x)$
is a $p_\ga$-form. It is easy to see that
\be
\delta R^\ga = (-1)^{p_\gb} \epsilon^\gb R^\gamma
\f{\p^2 F^\ga(\W)}{\p \W^\gamma \p \W^\gb}\,.
\ee
Hence, the unfolded equation (\ref{genunf})
is invariant under the gauge transformation (\ref{ggtr}).

The system (\ref{r0}), (\ref{freq}) describes
general linearized unfolded field equations, where $\W=(W_0,W_1)$ describes
background geometry via $W_0$ and dynamical fluctuations via $W_1$.
In the case where $F^\Omega (0, W_1)=0$ (\ie $W_0$ enters $F(W_0,W_1)$
at least linearly), which is considered in this paper,
differential forms of various degrees contained in $W_1$ are valued
in some  $g$-modules so that $d$ along with the terms linear in
$W_0$ in (\ref{Rco}) provide covariant derivatives of
$g$ in the respective modules. The terms of higher orders in
$W_0$, that cannot be removed by a field redefinition, describe
$g$-cohomology with the coefficients in the respective modules (see,
e.g. \cite{33} for more detail and references). As is well-known
\cite{Ann,33},  the
equations (\ref{freq}) can describe a field theory with infinite
number of degrees of freedom provided that
zero-forms contained in $W_1$ are valued in infinite dimensional
$g$-modules. In fact, the presence of infinite dimensional
modules is also crucial for the existence of a nontrivial $g$-cohomology
for  semisimple $g$, represented in this construction by
the part of $F^\Omega(W_0,W_1)$ of higher orders of $W_0$.

{}For the unfolded system (\ref{r0}), (\ref{freq}) with
$\W=(W_0\,,W_1)$ the transformation law (\ref{ggtr}) gives two types of symmetries.
The one associated to $W_1$ is Abelian because
$F(W_0\,,W_1)$ is linear in $W_1$ at the linearized level.
This is the Abelian gauge symmetry of free fields that generalizes
the transformation law (\ref{gtr}) to  $F(W_0\,,W_1)$
nonlinear in $W_0$.

The other one associated to $W_0$ is
\be\label{w0tr}
\delta W_0^A(x)=\D_0 \epsilon^A(x)
\ee
\be \delta W_1^\Omega(x)= (\epsilon (x)\cdot W_1(x))^\Omega\,,
\ee
where
\be \label{gltr1} (\epsilon \cdot W_1)^\Omega = -\epsilon^A
\frac{\p F^\Omega(W_0, W_1)}{\p W^A_0}\,.
\ee
This induces
nonAbelian global symmetries of the linear (free) system
(\ref{freq}). Indeed, having fixed a
background connection $W^A_0(x)$, we have to demand $\delta
W_0^A(x)=0$ which implies by virtue of (\ref{w0tr})
that the leftover global symmetry is
described by the global symmetry parameter $\epsilon_0^A (x)$ that
verifies
\be \label{d0} \D_0 \epsilon_0^A(x)=0\,. \ee
Since $\D^2_0 =0$, these equations are consistent, admitting a
solution that can be reconstructed from free values
$\epsilon^A(x_0)$ at any point $x_0$. Of course, this is a local
statement that can be obstructed by
 additional topological (\ie boundary) conditions. Assuming that this
does not happen (\ie that a manifold, where $W^A_0(x)$ is defined, is
indeed $g$--invariant) we conclude that the connections $W_1^\Omega$
and curvatures $R_1^\Omega$ transform covariantly \be \delta
W_1^\Omega(x)= (\epsilon_0 (x)\cdot W_1(x))^\Omega\q \delta R_1^\Omega(x)=
(\epsilon_0 (x)\cdot R_1(x))^\Omega\,
\ee
under  the global $g$--transformations. Hence, the equation
(\ref{freq}) is $g$ invariant.
Note that due to the  cohomological terms of higher orders in $W_0$,
the global $g$--transformation laws of the
$p$-form connections with $p>0$ acquire  via
(\ref{gltr1}) additional contributions that mix forms of
different degrees. (For  particular examples see Subsection \ref{stue}
and \cite{33}.)

In the particular case of global conformal symmetry
important for the further analysis it is elementary to solve
the condition (\ref{d0})
in the Cartesian coordinate system (\ref{mink}) to obtain
\bee \label{glsec}
\tilde{\epsilon}_{0\,a}(x) &=& \tilde{\varepsilon}_{a}\,,\\
\label{gldil}
\epsilon_0(x) &=& \varepsilon - x^a\tilde{\varepsilon}_{a}\,,\\
\label{gllor} \epsilon_0^{ab}(x) &=& \varepsilon^{ab} +
x^a\tilde{\varepsilon}^{b} -
x^b\tilde{\varepsilon}^{a}\,,\\
\label{gltr} {\epsilon}_0^a (x) &=& {\varepsilon}^{a} + x^b
\varepsilon^{ab} -x^a \epsilon +x^a x^b\tilde{\varepsilon}_{b}
-\half x^2 \tilde{\varepsilon}^{a} \,, \eee
where $\varepsilon_{a}$,
$\varepsilon$, $\varepsilon^{ab}$ and $\tilde{\varepsilon}^{a}$ are
arbitrary $x$--independent parameters of the global conformal
symmetry transformations which appear as integration constants of
the equations (\ref{d0}) and correspond, respectively, to
translations, dilatations, Lorentz transformations and special
conformal transformations. Let us stress  again that with this
choice of the conformal symmetry parameters, the Cartesian
vacuum connection (\ref{mink}) remains invariant under the conformal gauge
transformations (\ref{w0tr}).

Let $\cC^\go$ denote the subset of zero-forms among $W^\Omega_1$.
Since   in the sector of zero-forms $F^\omega(W_0, \cC)$ is
a one-form, it is linear in $W_0$ (recall that $W_0$-independent
terms are not allowed in our consideration).
Hence, $\cC^\go$ span some $g$-module $\cC$.
A distinguishing property of zero-forms is that they
have no associated gauge parameters and hence the inhomogeneous
(\ie the first) term in the transformation law (\ref{ggtr}).
As a result, at the linearized level, they describe gauge
invariant combinations of derivatives of dynamical fields.
Indeed, at least some of the zero-forms $\cC^\go$ turn out to
be expressed in terms of the gauge fields described by the $p$-form gauge
connections $W^\Omega_1$ with $p>0$ via the cohomological terms in (\ref{Rco})
that are nonlinear in $W_0$. Those, that are not expressed via derivatives of the
gauge fields by this mechanism describe non-gauge fields like, for
instance, scalar and spinor.  From this consideration it follows that
the zero-form module $\cC$ describes the space of gauge invariant
physical states. As such it is closely related to the module of
single-particle states in the corresponding field theory \cite{BHS}.

\subsection{Dynamical content via $\gs_-$ cohomology}
\label{sigma-}

We use the following terminology. The fields $W_1^\Omega $
contain {\it dynamical fields} $\phi^{dyn}$, {\it
auxiliary fields} $\phi^{aux}_n$  and {\it Stueckelberg fields} $\phi^{st}$.
As explained in more detail below, auxiliary fields  $\phi^{aux}_n$ are
expressed via
space-time derivatives of order up to $n$ of $\phi^{dyn}$ by appropriate
constraints. Stueckelberg fields $\phi^{st}$ are pure gauge, \ie they can be
gauge fixed to zero by algebraic Stueckelberg shift symmetries.
Leftover fields, that are neither Stueckelberg nor auxiliary, are dynamical.

The $(p-1)$--form gauge parameters describe  {\it differential gauge
symmetries} for the dynamical fields and algebraic {\it
Stueckelberg} (shift) symmetries that compensate the Stueckelberg
fields. The parameters (\ref{hgtr})
$\epsilon_i(x)$  with $i< p-1$, which exist for $p>1$, govern
the degeneracy of gauge transformations, \ie gauge symmetries for gauge symmetries.

The gauge invariant curvatures contain components of different types.
Some can
be set to zero by imposing {\it constraints} that express
algebraically auxiliary fields in terms of derivatives of the
dynamical fields. Some other are zero by virtue of  Bianchi identities
applied to the constraints. The leftover components, that remain
nonzero upon substitution of the expressions for auxiliary fields in
terms of derivatives of the dynamical ones, describe gauge invariant
combinations of derivatives of the dynamical fields. Among them, a
distinguished role have  {\it ground field strengths} that result
from the application of certain gauge invariant differential
operators to the dynamical fields and are such that all other gauge
invariant differential operators result from the ground ones by
further differentiations. In other words, all gauge invariant
combinations of derivatives of the dynamical fields are the ground field strengths
or their derivatives. A set of all gauge invariant combinations of
derivatives of the dynamical fields forms an infinite dimensional $g$--module
$\cC$ that fully characterizes the gauge invariant pattern of the
system. It is called Weyl module.

As an example, consider  Poincar\`{e} or $AdS$ gravity, where
$g=iso(1,d-1)$ or $o(d-1,2)$, respectively. The fields $W_1^\Omega$
take values in the adjoint representation of $g$, describing the
weak field deviations of the full connection $W^\Omega=
W_0^\Omega+W_1^\Omega$ from the vacuum one $W_0^\Omega$. They therefore
contain vielbein one-form $e^a$ and Lorentz connection $\go^{ab}$.
Here $\go^{ab}$ is the auxiliary field while $e^a$ contains the
dynamical field, which is the symmetric part of $e_{\un,a}$, \ie the
linearized fluctuation of the metric, and the Stueckelberg field,
which is the antisymmetric part of $e_{\un,a}$.
The latter is pure
gauge due to the (linearized) local Lorentz symmetry.

There are two types of gauge symmetries in this example. One is
local Lorentz symmetry with the gauge parameters $\epsilon^{ab}$,
which is Stueckelberg. Another one is the ``translation" gauge
symmetry with the vector gauge parameter $\epsilon^a$. This is the
true differential symmetry that describes linearized
diffeomorphisms.

The components $R_1^a$ of the gauge invariant curvatures, associated
to translations, can be set to zero by imposing the
zero--torsion constraint $R_1^a=0$ that expresses Lorentz connection
via the first derivatives of the vielbein. With this substitution,
$R_1^{\un\um}$ (with the convention $R_1^{ab} = e_{0\un}^a e_{0 \um}^b
R_1^{\un\um}$) is the linearized Riemann tensor two-form. The Bianchi
identities applied to the zero-torsion constraint imply
the familiar linearized cyclic identity of the Riemann tensor
\be e_0^a
\wedge R_1{}_a{}^b =0\,.
\ee
Other components of $R_1^{\um\un}$
remain algebraically independent. These include the linearized Ricci
tensor $\mathbf{R}_{\un\um}$ and the linearized Weyl tensor
$C_{\un\um,\uk\ul}$. They represent all gauge invariant combinations
of the metric that contain two derivatives, \ie the full set of
gauge invariant ground field strengths. Setting $\mathbf{R}_{\un\um}=0$
imposes the linearized Einstein (\ie spin two) equations. Setting
$C_{\un\um,\uk\ul}=0$ is the condition that the metric is
conformally flat (in the linearized approximation, that its
traceless part is zero). Imposing both conditions, \ie that the
linearized Riemann tensor is zero, implies that the linearized
metric is pure gauge.

Pattern of a general $g$--invariant system is  encoded in the
so-called $\sigma_-$ cohomology \cite{sigma} (see also
\cite{BHS,solv}) which is a perturbative concept that emerges in the
linearized analysis. To apply this machinery, the following
conditions have to be satisfied. First, a space $V$, where fields
$W_1^\Omega $ take their values, should be endowed with some grading
$G$ such that its spectrum is bounded from below. Usually $G$
counts a rank of a tensor (equivalently, a power of an appropriate
generating polynomial) and eventually counts the order $k-l$ of
highest space-time derivatives of degree $l$ dynamical fields
$\phi^{dyn}$ contained in the degree $k$ auxiliary fields
$\phi^{aux}$. In the example of gravity we set $G(A^n) = 1$ and
$G(B^{mn})=2$ for elements $A^n$ and $B^{nm}$ of the adjoint
representation of $g=iso(1,d-1)$ or $o(d-1,2)$.

Suppose that the background covariant derivative admits the
decomposition
\be \label{d0s} \D_0 = \D_{00} +\sigma_- +\sigma_+\,, \ee
where $ [G\,,\sigma_- ] = - \sigma_-\,, $ $ [G\,,\D_{00} ]= 0 $ and
$\sigma_+$ is a sum of some operators of positive grade. From
(\ref{d02}) it follows that $ \sigma_-^2 = 0\,. $ The standard
choice of $\sigma_-$ is such that it is the part of the covariant
derivative associated to the vielbein, that decreases the grading
$G$ by one. For instance, in the example of gravity \be
\label{adsgr} \sigma_- (A^a ) =0\q \sigma_- (B)^a = e_b B^{ba}\,.
\ee
(Recall that $\sigma_-$ maps a space of grade $G$ to the space
of grade $G-1$.) The grading $G$ should not be
confused with the grading that counts a degree of
a differential form. Since, in this paper, $\gs_-$ is a part of the
covariant derivative, it has the form degree one. Note that the more
general situation (\ref{Rco}) where $\sigma_-$ mixes differential forms of
different degrees is also of interest (for more detail see e.g. \cite{solv,33}).

Provided that $\sigma_-$ acts vertically ({\it i.e.,} does not
differentiate $x^\un$), the cohomology of $\sigma_-$ determines the
dynamical content of the dynamical system at hand.  Namely, as shown
in \cite{sigma}, for a $p$-form $W_1^\Omega$  valued in a vector
space $V$, $H^{p+1} (\sigma_- ,V)$, $H^{p} (\sigma_- ,V)$ and
$H^{p-1} (\sigma_- ,V)$ characterize, respectively, ground field
strengths (that can be interpreted as left hand sides of possible
gauge invariant field equations),
dynamical fields, and differential gauge symmetries
encoded by the curvatures (\ref{R1}), gauge fields $W_1$ and
transformation laws (\ref{gtr}).  Let us note that $H^{k}
(\sigma_- ,V)$ with $k>p+1$ describe so called syzygies\footnote{Syzygies
$H^{p+2} (\sigma_- ,V)$ describe differential relations (Bianchi
identities) between the ground differential operators, $H^{p+3}
(\sigma_- ,V)$ describe differential relations between the latter
differential relations, {\it etc}.} of the field equations \cite{33} while
$H^{k} (\sigma_- ,V)$ with $k<p-1$ describe differential gauge
symmetries for gauge symmetries (\ref{hgtr}) \cite{skvor}.

The meaning of this statement is simple.

From the level-by-level analysis of the linearized curvature $R_1$
(\ref{R1}) it follows that all fields that do not belong to
$Ker\,\gs_-$ can be expressed via derivatives of some lower grade
fields by setting appropriate components of the linearized curvature
$R_1$ to zero. Hence these are auxiliary fields. Those that are
$\gs_-$ exact can be gauged away by the Stueckelberg part of the
gauge transformation (\ref{gtr}) associated to the $\gs_-$ part
of $\D$ in (\ref{d0s}). The fields that remain belong to the
cohomology of $\gs_-$. These are dynamical fields.

Dynamical content of the gauge transformations and field
equations can be analyzed analogously. For example, suppose that it
is possible to impose constraints $R_{1\,i}=0$ on the auxiliary fields
imposing no restrictions on the dynamical fields, where
the curvatures $R_{1\,i}$ have grades  $i_{min}\leq i\leq k$.
{}From the
Bianchi identity (\ref{bi}) if follows then that $\gs_- R_{1\,k+1}=0$,
\ie $R_{1\,k+1}$ is $\gs_-$ closed. If the cohomology $H^{p+1}$ is zero
in the grade $k+1$ sector then the $R_{1\,k+1}=\gs_- (\chi_{k+2})$ for
some $\chi_{k+2}$. Since $W_{1\,k+2}$ also enters $R_{1\,k+1}$ via $\gs_-
(W_{1\,k+2})$, it can be adjusted to cancel $\chi_{k+2}$, thus
achieving $R_{1\,k+1}=0$. The latter condition becomes a combination of
identities that follow from the previously imposed constraints along with
the new one that expresses $W_{1\,k+2}$ via derivatives of the fields of lower
grades.  If, on the other hand, a nontrivial cohomology
$H^{p+1}$ appears on the level $k$, it is impossible to achieve that
$R_{1\,k}=0$ without imposing further differential equations on
$W_{1\,k+1}$ and hence on  the dynamical fields through which $W_{1\,k+1}$
has been already expressed. Equivalently, those components of $R_1$
that belong to $H^{p+1}$ represent gauge invariant ground field
strengths built from derivatives of the dynamical fields.
Clearly, if the dynamical fields have grade $l$, the resulting level
$k$ field strengths contain $k+1 -l$ their derivatives.
If a system is such that $H^{p+1}$ is zero, no
gauge invariant field equations are imposed by the equations $R_1=0$
and the system is called {\it off-shell}.

Since $\gs_-$ usually originates from the part of the covariant
derivative of  a space-time symmetry algebra that contains
vielbein, the nondegeneracy of the latter implies
that a maximal possible number of field components are expressed
via space-time derivatives of the dynamical fields.

\subsection{Conformal gauge fields as Fock vectors}
\label{Fock}

It is convenient to describe tensors as elements of an appropriate
Fock space. Consider a set of oscillators $a^{\dagger{}Ai}\,,
a^B_j$\,, where $1\leq i,j\leq h$ for  some $h\geq 1$, that satisfy
the commutation relations \be [a_i^A\,,a^{\dagger{}Bj}]= \delta_i^j
\eta^{AB}\q [a_i^A\,,a_j^{B}]=0\q
[a^{\dagger{}Ai}\,,a^{\dagger{}Bj}]=0\,. \ee

The bilinears
\be \label{t2} T^{AB} =\sum_i \left (a^{\dagger{}Ai}
a^B_i - a^{\dagger{}Bi} a^A_i \right )
\ee
satisfy the $o(d,2)$
commutation relations (\ref{d2c}).

The bilinears
\be \label{tau} \tau^{ij} = a^{\dagger{}Ai}
a^{\dagger{}Bj}\eta_{AB}\q \tau^i{}_j = \half \{
a^{\dagger{}Ai}\,,a^B_j\} \eta_{AB}\q \tau_{ij} = a^A_i
a^B_j\eta_{AB}\,
\ee
are generators of $sp(2h)$ with the following
nonzero commutation relations
\be
[\tau^i{}_j\,,\tau^k{}_l]=\delta_j^k\tau^i{}_l-\delta^i_l
\tau^k{}_j\q \ee \be [\tau^i{}_j\,,\tau{}_{kl}]=-\delta^i_k
\tau_{jl} - \delta^i_l \tau_{jk}\q
[\tau^i{}_j\,,\tau^{kl}]=\delta^k_j \tau{}^{il} + \delta^l_j
\tau^{ik}\q \ee \be [\tau_{ij}\,,\tau^{kl}] =
\delta_j^k\tau^l{}_i +\delta_i^k\tau^l{}_j + \delta_j^l\tau^k{}_i
+\delta_i^l\tau^k{}_j\,.
\ee

The $o(d,2)$ generators $T^{AB}$ rotate $o(d,2)$
vector indices $A,B$ with no effect on the $sp(2h)$ indices $i,j$
while the $sp(2h)$ generators $\tau$ act on the indices
$i,j$ with no effect on the vector indices $A,B$.
The two mutually commuting algebras $o(d,2)$ and $sp(2h)$  form
 so-called Howe dual pair \cite{Howe}.

Consider a Fock space $F$ spanned by vectors
\be \label{expan}
|\Phi\rangle = \sum_{L_1\geq 0,L_2\geq 0,\ldots }\frac{1}{\sqrt{L_1!
L_2 !\ldots}} \Phi^{A^1_1\ldots A^1_{L_1}\,, A^2_1\ldots
A^2_{L_2}\ldots} a^{\dagger{}1}_{A^1_1}\ldots
a^{\dagger{}1}_{A^1_{L_1} } a^{\dagger{}2}_{A^2_1}\ldots
a^{\dagger{}2}_{A^2_{L_2}}\ldots|0\rangle
\ee
generated from  the Fock vacuum
$|0\rangle$ that satisfies \be a_{i}^A |0\rangle =0\,.
\ee
{}From (\ref{t2}) it follows that homogeneous polynomials
$f(a^\dagger{})|0\rangle $ of different degrees form
finite dimensional $o(d,2)$--submodules of $V$. To single out
irreducible submodules one has to impose further conditions built from
the $sp(2h)$ generators that commute to the $o(d,2)$. These are the
$sp(2h)$ highest weight conditions
\be \label{tr} \tau_{ij}
|\Phi\rangle =0\,, \ee \be \label{asym} \tau^i{}_j |\Phi\rangle =0
\qquad j>i\,, \ee \be \label{length} \tau^i{}_i |\Phi\rangle= L_i
|\Phi\rangle \ee (no summation over $i$), where $L_i$ are some
non-negative integers such that
\be
\label{Lij}
L_j \geq  L_i\quad \mbox{for}\quad i>j\,.
\ee
Clearly, the condition (\ref{tr}) implies that the tensors in the
expansion (\ref{expan}) are traceless. The condition (\ref{asym})
implies that the symmetrization of all indices contracted with the
oscillators
 $a^{\dagger{}iA}$ for some $i$ with any index contracted with an oscillator
 $a^{\dagger{}jA}$ with $j>i$ gives zero. This is the Young antisymmetrization
 condition in the so-called symmetric basis (indices contracted with the
 bosonic oscillators $a^{\dagger{}iA}$ for a given $i$ are automatically
 symmetrized). Finally, the condition (\ref{length}) determines
 a number of indices in the manifestly symmetrized groups.
 Such a tensor is depicted by the Young diagram composed of
 $h=H_1$ rows of lengths $L_i$
\be\label{Yd2}
\YoungGeneralized
\vspace{0.3cm}
\ee
 Note that Young diagrams that do not satisfy (\ref{Lij})
are not considered
because in this case the conditions (\ref{asym}) and (\ref{length})
admit only zero  solution. Another restriction specific for the
modules of the orthogonal algebra, that
results from the tracelessness conditions, is that
traceless Young diagrams of $o(k,l)$ can be nonzero provided that the
heights $H_1$ and $H_2$ of its first (and hence any) two columns
satisfy
\be
\label{ham}
H_1+H_2\leq k+l\,.
\ee
Indeed, this is easily seen by  the dualization of
the two columns, that is by the contraction of indices
with two epsilon symbols
$$\varepsilon_{A_1\ldots A_{k+l}}
\varepsilon_{B_1\ldots B_{k+l}} a^{A_1}_1\ldots a^{A_{h_1}}_{h_1}
a^{B_1}_1\ldots a^{B_{h_2}}_{h_2}
$$
which, by virtue of the determinant formula for two epsilon symbols,
 turns out to be proportional to the operator $\tau_{ij}$
if (\ref{ham}) is not respected, because  at least one
 pair of indices of the two chosen columns will be contracted.
(See, e.g., \cite{ham}.)

The Young diagram (\ref{Yd2}) will be denoted  $Y(\bL|o(d,2))$
where
\be
\bL = (L_1,L_2,\ldots , L_h )\,.
\ee
The space of such tensors will be denoted $F^\bL_{o(d,2)}$.
 To single out an irreducible
$o(d,2)$ tensor it remains to impose the (anti-)selfduality
condition, that is possible for $d=2\, mod\, 4$ and $h=d/2+1$.
We will not consider (anti-)selfdual tensors in this paper, however.

The space $\Lambda$ of differential forms in $d$ dimensions
is also convenient to realize as a Fock module. To this end we
introduce fermionic
annihilation and creation operators $\xi_a$ and $\xi^{\dagger a}$,
respectively, that carry Lorentz indices and have anticommutation
relations \be \{\xi_a\,,\xi^{\dagger b}\}=\delta^b_a\q
\{\xi_a\,,\xi_b\}=0\q \{\xi^{\dagger a}\,,\xi^{\dagger b}\}=0\,.
\ee
We identify $\xi^{\dagger\,a}$ with tangent components of space-time
differentials $dx^\un$ via
\be \xi^{\dagger\,a}=dx^\un h_\un^a\,.
\ee
Then $\Lambda$ is realized as the space of states
\be
|\go \rangle =
\sum_{p=0}^d  dx^{\un_1}\ldots dx^{\un_p} \go_{\un_1\ldots
\un_p}|0\rangle = \sum_{p=0}^d
\xi^{\dagger\,a_1}\ldots \xi^{\dagger\, a_p} \go_{a_1\ldots
a_p}|0\rangle\,, \ee
where the Fock vacuum is defined by
\be
\xi_{ a} |0\rangle =0\,.
\ee
We shall often omit the wedge symbol
because the wedge product of differential forms is nothing but the
product in the Grassmann subalgebra generated by $\xi^{\dagger a}$ of the Clifford
algebra generated by $\xi_a$ and $\xi^{\dagger b}$. $p$ forms span
the subspaces $\Lambda^p$ of degree $p$--homogeneous
polynomials in $\Lambda$, \ie
 \be \label{form} |\omega \rangle \in
\Lambda^p \,:\qquad \xi^{\dagger a}\xi_a  |\omega \rangle = p
|\omega \rangle\,.
\ee

The full Fock space $\cF$ is \be \cF = F\otimes \Lambda\,. \ee Its
subspace $\cF_{o(d,2)}^\bL$,
 that satisfies the conditions (\ref{tr})-(\ref{length}),
is \be \label{L} \cF_{o(d,2)}^\bL = F_{o(d,2)}^\bL \otimes
\Lambda\,. \ee The subspace associated to $p$--forms is \be
\label{Lp} \cF_{o(d,2)}^{\bL, p} = F_{o(d,2)}^\bL \otimes
\Lambda^p\,. \ee

Let
\be \label{expan1} |\Omega\rangle =
\sum_{p\geq 0;L_i\geq 0} \frac{1}{\sqrt{\,L_1! L_2 !\ldots}}
\Omega_{a_1\ldots a_p}^{A^1_1\ldots A^1_{L_1}\,, A^2_1\ldots
A^2_{L_2}\ldots} \xi^{\dagger\, a_1}\ldots \xi^{\dagger\,a_p}
a^{\dagger{}1}_{A^1_1}\ldots a^{\dagger{}1}_{A^1_{L_1} }
a^{\dagger{}2}_{A^2_1}\ldots
a^{\dagger{}2}_{A^2_{L_2}}\ldots|0\rangle\,.
\ee
We endow the Fock space $\cF$ with
the scalar product that pairs $p$-forms with $p$-forms
\be
\label{stin} \langle \Phi |\Omega \rangle = \sum_{p\geq 0;L_i\geq
0}{p\,!\,}\overline{\Phi}^{a_1\ldots a_p}_{A^1_1\ldots
A^1_{L_1}\,, A^2_1\ldots A^2_{L_2},\ldots} \Omega_{a_p\ldots
a_1}^{A^1_1\ldots A^1_{L_1}\,, A^2_1\ldots A^2_{L_2},\ldots}\,.
\ee
Note that
\be \langle 0 |\xi^{\dagger\,a} =0\q \langle
0 |a^{\dagger\,i}_A =0\q \langle 0 |0\rangle =1\,.
\ee

\section{$\sigma_-$ cohomology analysis}
\label{Sigma}
\subsection{Operator $\sigma_-$}
\label{sigcon}
Let a $p$--form $|\Omega(x)\rangle$ be realized as a Fock vector
(\ref{expan1}) that satisfies the
conditions (\ref{tr})-(\ref{length}) and (\ref{form}).
In other words, it is a section of the vector fiber bundle
over $d$-dimensional Minkowski space $M^d$
($x^\un $ are any local coordinates of $M^d$)
with the fibers $F_{o(d,2)}^{\bL,p}$.

For any fixed conformal invariant background described by a flat
connection $W_0$ (\ref{r0}), the linearized HS curvature $R_1$ is
\be
|R_1(x)\rangle =
\D_0|\Omega(x)\rangle\q \D_0 = d+\half W_{0\,AB} T^{AB}\q
d=dx^\un\f{\p}{\p
x^\un}=\xi^{\dagger\,a} h^\un_a \f{\p}{\p x^\un}\,.
\ee

We define $\sigma_-$ as the part of covariant derivative
associated to the translation generator of the conformal algebra,
\be \label{s-} \sigma_- = \xi^\dagger_a T^{-a}\,.
\ee
Let us stress that $\cF_{o(d,2)}^\bL$ is invariant space of $\gs_-$  for any
$\bL$.

The grading operator $G$ of Section \ref{sigma-} is the
dilatation generator $D=T^{+-}$
\be \label{dil}
G=D = n_+ -n_-\,,
\ee
where
\be \label{npm}
n_+ = a^{\dagger{}+ i} a^-_i\,\q n_- = a^{\dagger{}- i} a^+_i.
\ee
$G$ counts the difference between the number
of indices that take values $-$ and $+$ in the coefficients
$\Omega_{a_1\ldots a_p}^{A^1_1\ldots A^1_{L_1}\,, A^2_1\ldots
A^2_{L_2}\ldots}(x)$ in Eq.~(\ref{expan1}) (recall that $A^\pm
=A_\mp$). Note that $0 \leq n_\pm \leq L_1$
 because the Young properties imply that symmetrization over
any $L_1+1$ indices gives zero.

Generally, the grading $G$ may or may not be
induced by some grading of the symmetry algebra $g$. For example, in
the case of $AdS_d$ gravity  with $g=o(d-1,2)$, the grading
(\ref{adsgr}) is not a grading of $g$. (From the conformal algebra
perspective this is so because the $AdS$ translation generators are
represented by a mixture of conformal translation generators and
special conformal generators that carry different conformal
dimensions.) In this case, $\sigma_-$ acts in $V$ but does not belong
to the representation of $g$. Depending on whether $\sigma_-$ belongs to
$g$ or not, $\sigma_-$ cohomology is related to the Lie algebra
cohomology  (as in the conformal case) or not (as in the $AdS$ case).

\subsection{Homotopy operator}
\label{homop}
\subsubsection{Generalities}
\label{gen}
Let $\V= \sum_{p=-\infty}^{\infty} \oplus \V^p $ where $\V^p$ is
some set of finite dimensional linear spaces.
Let $\gs$ be a grade one nilpotent operator
\be
\label{1} \gs (\V^p) \subset \V^{p+1}\q \gs^2 =0\,.
\ee

Cohomology $H^p(\sigma, \V)$ is the quotient space
\be
H^p(\sigma, \V) =  \frac{ Ker \,\gs \cap \V^p }{ Im \,\gs \cap\V^p }  \,.
\ee
Roughly speaking, $H^p(\sigma, \V)$ consists of elements
of $\V^p$, that are $\sigma$ closed but not exact.

Let $\sigma^\dagger$ be a grade $-1$ operator that squares to zero, \ie
\be \label{-1} \sigma^\dagger (\V^p) \in
\V^{p-1}\q (\sigma^\dagger)^2=0\,.
\ee
Then, from (\ref{1}) and
(\ref{-1}) it follows that
 the {\it homotopy operator}
\be \cH =\{\sigma\,,\sigma^\dagger\}
\ee
commutes both with
$\sigma$ and with $\sigma^\dagger$
\be \label{comss}
[\sigma\,, \cH ] =0\q [\sigma^\dagger\,, \cH ]=0\,
\ee and
\be
\cH (\V^p ) \in \V^p\q \forall p\,.
\ee

An efficient standard  cohomology
computation tool (see e.g. \cite{HT}) is provided by the\\
 \noindent
\emph{\textbf{Homotopy Lemma }\\
If $\cH$ is diagonalizable in $\V$, then $H=\sum_p\oplus  H^p(\sigma,
\V)\subset Ker \, \cH $.}\\ \\
 Indeed, let $v$ be a
$\sigma$--closed eigenvector of $\cH$
\be
\cH v = \lambda v\q \sigma v = 0
\ee
with $\lambda\neq 0$. Then $v$ is
$\sigma$--exact because \be v=\lambda^{-1}\cH\, v = \sigma
\chi\q\chi=\lambda^{-1}\sigma^\dagger  v\, \qquad\qquad\qquad\blacksquare
\ee

In other words, only those elements that belong to the subspace
spanned by eigenvectors of $\cH$ with zero eigenvalue can belong to
$H$.

Now we are in a position to apply Homotopy Lemma  to the operator
$\gs=\gs_-$  with $\V = \cF_{o(d,2)}^\bL$ (\ref{L}) and $\V^p
= \cF_{o(d,2)}^{\bL,p}$ (\ref{Lp}). (Let us stress that the operator
$\gs_-$ (\ref{s-}) increases a form degree, with respect to which the
cohomology $H^p$ is defined, but decreases the grading
in $o(d,2)$.) This will allow us to compute
$Ker\, \cH$ and, hence, $H^p(\sigma_-, \V)$ for all $p$. The method
provides a nice application of supersymmetry and turns out to be
surprisingly simple and efficient. Hence, we explain it in some
detail.

\subsubsection{Evaluation of $\cH$}
\label{evalh}

Let
 \be \sigma_-^\dagger
=-\xi_a T^{+a}\,. \ee Using (\ref{d2c}),  (\ref{t2}), (\ref{s-}),
(\ref{dil}) and  (\ref{npm})
 it is easy to obtain
\be
\label{H1}
-2\{\sigma_-\,,\sigma_-^\dagger\}= -2\cH = \{T^{-a}\,,T^+_a\} -
T^{ab}T^F_{ab} +D[\xi^\dagger_a\,,\xi^{a}]\,,
\ee
where $T^{ab}$
and $T^F_{ab}$ are the Lorentz generators that act, respectively, on
the bosonic and fermionic oscillators,
 \ie $T^F_{ab}$ act on the indices of differential forms
\be T^F_{ab} = \half \left ( [\xi_{a}\,,\xi^\dagger_{b}]-
[\xi_{b}\,,\xi^\dagger_{a}]\right )\,. \ee
Let \be
T^L_{ab}=T_{ab}+T^F_{ab}\,
\ee
be the total Lorentz generator that
rotates all Lorentz indices. Then Eq.~(\ref{H1}) gives
\be \label{sigma} \cH = \f{1}{4}\Big (T^{L\,ab}T^L_{ab} -T^{AB}
T_{AB} \Big ) -\half (\Delta+p)(\Delta+p -d)\,,
\ee
where $p$ is a
form degree and $\Delta$ is the conformal dimension
\be
\xi^\dagger_a
\xi^a |\Omega\rangle = p\, \Omega \q D\Omega = \Delta
|\Omega\rangle\,.
\ee
Hence, to evaluate $\cH$, it is enough to
evaluate the Casimir operators of $o(d,2)$ and $o(d-1,1)$ on their
arbitrary finite dimensional modules.

An elementary computation by using the fact that the Casimir
operators of $o(d,2)$ are expressed via those of the Howe dual
algebra $sp(2h)$, which in turn are easily computed using
(\ref{tr})-(\ref{length}), gives the following well-known result
\cite{BR}
\be
\label{symcas} T^{AB} T_{AB} =-2\sum_{i=1}^h L_i (L_i +d+2 -2i )\,,
\ee where summation is over all rows of the Young diagram.

Note that analogous computation in terms of fermionic oscillator
realization of Young diagrams, that makes antisymmetries manifest,
gives \be T^{AB} T_{AB} =2\sum_{i=1}^{L_1} H_i (H_i -d -2i )\,, \ee
where $i$ enumerates columns of the same Young diagram, that have
heights $H_i$  ($H_1=h$). Although it is  not
immediately obvious that these two expressions give the same result
for any Young diagram, the identity can be checked by induction.

It is sometimes convenient to represent a Young diagram as a combination
of rectangular horizontal blocks composed of rows of equal lengths.
In this representation, a Young diagram is characterized by the
lengths $L_\ga$ and heights $F_\ga$ of the $\ga^{th}$ blocks. Again,
\be L_\ga \geq L_\gb\qquad \mbox{for}\quad \ga < \gb\,. \ee In
particular, one can consider the case with $F_\ga=1$ that
corresponds to the standard description of a Young diagram in terms
of  rows. The maximal block decomposition of a Young diagram
is that with pairwise different lengths of all blocks,
\ie $L_\ga > L_\gb$ {for} $\ga < \gb$
\be\YoungGeneralDifBl
\ee

For any block decomposition, Eq.~(\ref{symcas}) gives \be
\label{symbl} T^{AB} T_{AB} =-2\sum_{\ga} L_\ga F_\ga (L_\ga +d+1
-2\sum_{\gb=1}^{\ga-1}F_\gb - F_\ga)\,. \ee

The Casimir operator  $T^{AB}T_{AB}$ in (\ref{sigma}) is determined by
the choice of the $o(d,2)$--module $\cF_{o(d,2)}^\bL$. The rest
terms in (\ref{sigma}) are diagonal on irreducible modules $V_u$
of the Lorentz subalgebra $o(d-1,1)\subset o(d,2)$ contained in the
tensor (\ref{expan1}), \ie
\be \label{pat} Res^{o(d,2)}_{o(d-1,1)}\cF_{o(d,2)}^\bL=\sum_u \oplus V_u\,.
\ee

Since indices of a differential $p$-form in (\ref{expan1}) are totally
antisymmetrized,  they are described by a column of height
$p$. Hence, the $o(d-1,1)$ pattern of
the $p$--form (\ref{expan1}) is
\bee \Omega = \sum_u \oplus \col \otimes V_u\,.
\eee
This can be worked out  with the help of the two simple facts
explained in Subsections \ref{TPD} and \ref{DR}.

\subsubsection{Tensor product with a differential form}
\label{TPD}

Let $V_u$ be described by some Young diagram $Y(l_i|o(d-1,1))$ and
\be
\label{pd}
p\leq [{d}/{2}]\q l_{h+1}=0\q
h\leq [{d}/{2}]\,.
\ee
 Then
\be \label{tencol}\!\!\!\!
 Y(\underbrace{1,1,\ldots 1,}_p 0,\ldots 0)\otimes Y(l_1,l_2\ldots )=
\sum_{\{\varepsilon_i\}} \oplus r(\varepsilon)\,Y(l_1+\varepsilon_1, l_2+\varepsilon_2,
\ldots )\q \varepsilon_i = 0 \quad\mbox{or}\,\,\, \pm 1 \ee at the
condition that
 \be
\sum_{i=1}^h \varepsilon^2_i \leq p\,
\ee
and the resulting Young diagram is admissible, \ie
the conditions (\ref{Lij}) and (\ref{ham}) are respected.
A positive integer $r(\varepsilon)$ is the multiplicity of
the diagram $Y(l_1+\varepsilon_1, l_2+\varepsilon_2,
\ldots )$ with a given set $\varepsilon_i$.

The meaning of this formula is as follows. In the tensor product
(\ref{tencol}), each index of the column $ Y(\underbrace{1,1,\ldots
1,}_p 0,\ldots 0)$, referred to as form indices,
 can be either added or subtracted
from the diagram $Y(l_1,l_2\ldots )$ (subtraction results from the
contraction of a pair of indices between the two factors in the
tensor product). Because indices in a row are symmetrized while
the tensor $ Y(\underbrace{1,1,\ldots 1,}_p
0,\ldots 0)$ is totally antisymmetric, no two form indices can be
added or subtracted to the same row. So, there are three options. Some of form indices are
added to some rows. These are labelled by $\varepsilon_i =1$. Some
other are subtracted from some other rows. These are labelled by
$\varepsilon_i =-1$. The third option is that one form index is
contracted to a row and then some other is added to the same row.
Such rows, as well as not involved ones, are labelled by
$\varepsilon_i =0$. If the third option occurs
in some $\half(p-\sum_{i=1}^h (\varepsilon^2_i))>0$ rows it leads to
the degeneracy $r(\varepsilon )>1$ . Fortunately, modules of this type will be shown not to
contribute to the cohomology of interest, that allows us to avoid
the computation of the multiplicities $r(\varepsilon)$. For our
purpose it is enough to use

\noindent
\emph{\textbf{Lemma 1}\\
For $p\leq d/2$,
any Young diagram with all form indices added, \ie $
\sum_i\varepsilon_i = p\,, $ appears once.}

\noindent
\emph{\textbf{Comment}\\
By dualization, in the case of $p\geq d/2$  the same
is true with $p$ replaced by $p^\prime=d-p$.}

\subsubsection{Dimension reduction}
\label{DR}

Now let us discuss the pattern of the decomposition (\ref{pat}). Let
an $o(d,2)$-module $F_{o(d,2)}^\bL$ be characterized by a Young
diagram with the maximal horizontal rectangular blocks of lengths
$L_\ga$ and heights $F_\ga$, \ie abusing notation,
$\bL =
((L_1,F_1),(L_2,F_2),\ldots )\,.
$

Let $d+2$ dimensions be decomposed
into $d+1$ dimensions plus one distinguished dimension
along a vector $V_A$. Clearly, all $o(d,1)$ tensors contained in the
original $o(d,2)$ tensor result from various
projections along $V_A$. These are described by those
$o(d,1)$ Young diagrams that result from the original $o(d,2)$ \YD
by cutting any number of cells in such a way that the resulting \YD
is admissible and no two cut cells belong to the same column because
projection along the same vector
$V_A$ is symmetric in all projected (\ie cut) indices. It
remains to note that the latter symmetry property implies that all cut
cells of any block can be moved to its bottom row. This
gives the following branching
\be \label{dimred} Res^{o(d,2)}_{o(d,1)} F_{o(d,2)}^\bL
\Longrightarrow \sum_{\tilde{L}_\ga}\oplus F_{o(d,1)}^{\bL(\tilde{L}_\ga)}
\ee
where
\be \bL(\tilde{L}_\ga) =  \Big( (L_\ga,F_\ga
-1),(\tilde{L}_\ga,1)\Big )\q L_{\ga+1}\leq \tilde{L}_\ga \leq
L_\ga\,,
\ee
\ie \bee \nn {\begin{picture}(600,150)(10,0)
\put(185,75){$\Longrightarrow$}

\YoungGeneralDifBll\quad \quad
  \quad\YGBR
\end{picture}}
\eee

\vspace{-8mm} As a result, we arrive at

\noindent
\emph{\textbf{Lemma 2}\\
Pattern of $Res^{o(d,2)}_{o(d,1)} F_{o(d,2)}^\bL$ results from cutting any
number of cells from the last row of each maximal block in such a way that
the resulting Young diagram  is admissible and no two cut cells belong to
the same column. Every diagram in this
list appears  once.}

The branching of $o(d,2)$--modules into $o(d-1,1)$--modules can be
obtained by the repetition of this procedure.

For our purpose, it is convenient to assume that the two types of cut indices are
$+$ and $-$. Clearly, the result of branching is insensitive to
whether the projected direction is time-like, space-like or light-like as
is the case for the $\pm$ directions. More precisely, we will
assume that irreducible Lorentz tensors carry $n_+$ lower indices
$+$ and $n_-$ lower indices $-$, \ie the oscillators to which these
indices are contracted carry $n_\pm$ upper indices $\pm$. (Abusing
notation, we use the same notation for the operators (\ref{npm}) and
their eigenvalues.) This means that for each $o(d-1,1)$ tensor
resulting from the double dimensional reduction of some $o(d,2)$ tensor,
its conformal weight $\Delta$ is
\be \label{delta} \Delta
=n_+ - n_-\,. \ee

It is straightforward to compute the homotopy operator $\cH$ on
any irreducible Lorentz submodule by using the formula (\ref{sigma})
and the formulae for the Casimir operators like (\ref{symcas}) or
(\ref{symbl}). Although $\cH$ is automatically diagonal in this basis,
for the first sight,  it may seem difficult to find
$Ker\, \cH $. The trick that allows us to solve the problem
is to use underlying supersymmetry.

\subsection{Cohomology and supersymmetry}
\label{sup} To observe supersymmetry it is enough to change
notations \be Q= \sigma_-\q Q^\dagger = \sigma_-^\dagger\,. \ee The
supergenerators $Q$ and $Q^\dagger$ are  conjugated with respect to
the conjugation that maps $a^B_i$ to $a^{\dagger i}_B$ and $\xi^a$
to $\xi^{\dagger a}$. In the
unitary case of $o(d+2)$ with positive-definite $\eta^{AB}$, the
scalar product on the Fock space $\cF$, that respects the involution
$\dagger$, is positive-definite. In this truly supersymmetric case,
the homotopy operator (Hamiltonian)
 \be
\cH = \{Q\,,Q^\dagger\}
 \ee
is non-negative, \ie all eigenvalues of $\cH$ are either positive or
zero.

The finite dimensional space $\cF^\bL_{o(d,2)} \subset \cF$
is invariant under the action of $Q$, $Q^\dagger$ and $\cH$. So,
$Ker\, \cH$ can be searched separately for each of these spaces.
The key observation, which will allow us to find easily $Ker\, \cH$
and then cohomology $H^p(\sigma_-,F^\bL_{o(d,2)})$, is

\noindent
\emph{\textbf{Lemma 3}\\
The homotopy operator $\cH$ is
non-negative on $\cF^\bL_{o(d,2)}$.}

This follows from the observation that the eigenvalues
(\ref{sigma}) of $\cH$ are insensitive to the signature of the
metric $\eta^{AB}$, depending only on the
type of the Young diagrams and  eigenvalues $\Delta$.
In the compact case of $o(d+2)$, where $\cH$ is non-negative,
the eigenvalues of $\Delta$  are still given by Eq. (\ref{delta}),
where $\pm$ components
correspond to the complex notation on a Euclidean two-plane: $A^\pm
=\f{1}{\sqrt{2}}( A^0\pm i A^d)$. Although in this case
the generators $T^{\pm A}$ are complex conjugated so that, strictly
speaking, each of them does not belong to  $o(d+2,\mathbb{R})$,
this does not affect the computation  in the
complex Hilbert space $\cF$ $\blacksquare$.

Before going into detail of the computation,
which is the subject of
Section \ref{comp}, let us note that,
using that the relevant spaces are finite dimensional,
in the unitary case of $o(d+2)$,
 a stronger version of the Homotopy Lemma of Section \ref{gen} holds:

\noindent
\emph{\textbf{Lemma 4}}\\
\be \label{fact} H(Q,\cF^\bL_{o(d+2)}) = Ker \cH\Big
|_{\cF^\bL_{o(d+2)}}\,. \ee
Proof: in the unitary
case $Ker\, \cH$ consists of supersymmetric states annihilated both
by $Q$ and by $Q^\dagger$ \be \langle a |\{Q\,,Q^\dagger\} |a\rangle
=0\quad \longrightarrow\quad Q|a\rangle=0\q Q^\dagger |a\rangle
=0\,.
\ee
Hence,  elements of $Ker\, \cH$ are
$Q$-closed. To show that $Ker
\Sigma$ does not contain $Q$-exact elements, suppose that
\be
 \cH |a\rangle =0\q |a\rangle = Q|b\rangle
\ee for some $|b\rangle$. Since $ [Q,\cH ]=0 $ ({\it cf}
(\ref{comss})),  the expansion of $|b\rangle$ in eigenvectors
of $\cH$ can only contain those with  zero eigenvalues, \ie
$|b\rangle\in Ker\,\cH$. But this implies that $Q|b\rangle =0$, \ie
every exact $|a\rangle\in Ker\,\cH $ is zero $\blacksquare$

Although, {\it a priori}, one has to be careful with the
extrapolation of Lemma 4 beyond the unitary case,  the following

\noindent
\emph{\textbf{Proposition}\\
Eq. (\ref{fact}) holds for  $o(d,2)$}\\
turns out to be true. This happens because  the signature of the
metric neither  affects  the cohomology computation nor obstructs
(anti-)selfduality conditions due to complexification of the
space of states $\cF$.

\subsection{Cohomology computation}
\label{comp}

First of all, we observe that the expression (\ref{sigma}) for
$\cH$ is symmetric under the Hodge duality supplemented by the
sign change  of  conformal dimension
\be
p\to d-p\q
\Delta \to -\Delta\,.
\ee
As a result, Proposition  of Section
\ref{sup} suggests that \be H^p(\gs_-, \cF^\bL_{o(d,2)} )
=H^{d-p}(\gs_-, \cF^\bL_{o(d,2)} )\,. \ee Since $\Delta$ changes
its sign under the exchange of indices $+$ and $-$,
representatives of $H^{d-p}(\gs_-, \cF^\bL_{o(d,2)})$ and
$H^p(\gs_-, \cF^\bL_{o(d,2)})$ are related by the same exchange.
We therefore consider the case $p\leq d/2$. Also it follows from
this symmetry that in the case of $p=d/2$ for even $d$ and
$\Delta\neq0$ the cohomology is doubled due to the exchange
between $+$ and $-$.

The idea of computation is to study the dependence of $\cH(\mu)$
on parameters $\mu$ that characterize various irreducible Lorentz
modules in the system. If, for some $\mu$,
we find that $\cH(\mu)>\cH(\mu^\prime)$ for some other possible
$\mu^\prime$, then $\cH(\mu)>0$ by Lemma 3. This simple
analysis allows us to rule out most of the possibilities, and easily
find $Ker\,\cH$. (The reader not interested in details
can go directly to the Theorem in the end of this section.)

Parameters that can vary include

\begin{itemize}
\item
The original $o(d,2)$ diagram $Y(\textbf{L}|o(d,2))$.
\item
The numbers $n_+$ and $n_-$ of indices associated to the Lorentz
invariant directions $+$ and $-$ and, in particular, the conformal
dimension (\ref{delta}). Different positions of cells $\+$ and $\mi$
also matter.
\item
Parameters $\varepsilon_i$ that characterize various types of
irreducible Lorentz tensors resulting from the tensor product with
the form indices in (\ref{tencol}), as well as the  form rank $p$.
\end{itemize}

It is important to note that this analysis is true only for those
Young diagrams that describe nonzero tensor modules, \ie satisfy
(\ref{Lij}) and (\ref{ham}).

Let the original $o(d,2)$ diagram  $\textbf{Y}$ be such that
$h_1\leq [\frac{d}{2}]+1$. This condition guarantees that the
restriction (\ref{ham}) holds and can always be reached by
dualization.

{}From (\ref{sigma}) we obtain

\noindent \emph{\textbf{Lemma 5}\\
a. For $p< [{d}/{2}]$ $Ker \cH$ has
$\Delta \leq 0$, \ie $n_- \geq n_+$. For $p >[{d}/{2}]$ $Ker \cH$
has $\Delta \geq 0$, \ie $n_- \leq n_+$.\\
b. For $p= [{d}/{2}]$ $Ker \cH$ is symmetric under the exchange of
 \+ and \mi.}\\
Proof: from (\ref{sigma}) it follows that the exchange of \+ and
\mi which changes $\Delta$ to $-\Delta$ only affects the
$\Delta$--dependent term in (\ref{sigma}) which is
$-\frac{1}{2}(\Delta^2 -(d-2p)\Delta)$. The latter is even in
$\Delta $ if $p= [{d}/{2}]$ and smaller (greater) for negative
(positive) than for positive (negative) $\Delta$ if $p< [{d}/{2}]$.
$\blacksquare$

\noindent
\emph{\textbf{Lemma 6}\\
Varying  various $\varepsilon_i$ in (\ref{tencol}),  $\cH$ is
minimized at
\be \label{eps}
\varepsilon_1=\ldots=\varepsilon_{p^\prime}=1\q
\varepsilon_{p^\prime+1}=\ldots= \varepsilon_h = 0\q p^\prime =
min\, (p,d-p)\,,
\ee
\ie all (dualized for $2p>d$) form indices are
added to the first $p^\prime$ rows.}\\
This follows from the observation that only the first term in
(\ref{sigma}) changes for different $\epsilon_i$ at given $p$ and
that from the formula (\ref{symcas}) applied to the $o(d-1,1)$
modules it follows that $\cH$ is minimized in the case (\ref{eps}).
$\blacksquare$

Let $F^{\bL\,rel}_{o(d,2)}\subset F^\bL_{o(d,2)}$ be spanned by
\emph{relevant} subspaces $V_v^{rel}$ \be F^{\bL\,rel}_{o(d,2)} =
\sum_v \oplus V_v^{rel} \ee that are described by those Young diagrams in
the Lorentz decomposition of $Y(\textbf{L}|o(d,2))$ that have no
Lorentz indices in the $([\frac{d}{2}]+1)^{th}$ row, \ie the lowest
row of $Y(\textbf{L}|o(d,2))$ is either zero or fully filled with \+
and \mi\, cells. Also we use notation
\be \label{oreg}
\Omega^{rel}=\sum_v \oplus \col\otimes V_v^{rel}\,.
\ee

The quotient space $V/V^{rel}$ is zero if the lowest row of
$Y(\textbf{L}|o(d,2))$ is zero \ie $H_1\leq [\frac{d}{2}]$. If $H_1
= [\frac{d}{2}]+1$, then $V/V^{rel}$ is spanned by those $V_u$ that
are described by the Lorentz Young diagrams that have one Lorentz
vector cell in the $([\frac{d}{2}]+1)^{th}$ row and still respect
the condition (\ref{ham}) (note that two Lorentz cells in the
$([\frac{d}{2}]+1)^{th}$ row is incompatible with (\ref{ham})).
This case, however, is ruled out by

\noindent
{\it\textbf{Lemma 7}\\
 $\ Ker \cH \subset \Omega^{rel}$.}\\
The proof follows from Lemma 6 along with the observation that
by formula (\ref{symcas}) applied to the Lorentz algebra $o(d-1,1)$
(\ie with $d+2$ replaced by $d$) the
contribution of the Lorentz cell in the $[\frac{d}{2}]+1$ row of the
Lorentz diagram to the first term in (\ref{sigma}) is zero for odd
$d$ and $\half$ for even $d$. Replacing this cell by \+ or \mi
therefore does not increase this contribution. Using Lemma 5 we
observe that in the case of  $p\leq [{d}/{2}]$, $\Delta \leq 0$
  ( $p\geq [{d}/{2}]$, $\Delta \geq 0$)
the replacement of the Lorentz cell in the $[\frac{d}{2}]+1$ row
by a \mi (\+) cell decreases the last term in (\ref{sigma}).
Hence $\ Ker \cH$ is minimized on $\Omega^{rel}$.$\blacksquare$

Thus, it remains to consider $\Omega^{rel}$. The following simple
observations  solve the problem.

\noindent \emph{
\textbf{Lemma 8}\\
$Ker\, \cH$ consists of those Young diagrams in $\Omega^{rel}$
that contain either  $\+$ cells or $\mi$ cells.}\\
Proof: firstly, we observe that
if there is a pair of \+ and \mi\, in some
column, its removal leads to the Lorentz diagram
that can be obtained  from the $o(d,2)$ Young diagram with the
removed two cells at the same positions.
In this case only the second term in Eq.~(\ref{sigma}) changes and from
Eq.~(\ref{symcas}) it follows that the resulting diagram has smaller
$\cH$.

\bee 
{    \begin{picture}(200,100)(-10,110){
\put(0,157){$\left\{\rule{0pt}{53pt} \right.$}
\put(-40,160){ {$[d/2]\!\!+\!\!1$}}
 \put(0,190) { \YoungBlock{17}{2}
}
\multiput(117,191)(10,0){04}{\pls}%
\multiput(157,191)(10,0){02}{\mns}%
\multiput(157,201)(10,0){02}{\pls}%

\put(0,140) { \YoungBlock{11}{5} }
\multiput(87,141)(10,0){03}{\mns}%
\multiput(87,151)(10,0){03}{\pls}%

\put(0,110){ \YoungBlock{8}{3} }
\multiput(07,111)(10,0){01}{\pls}%
\multiput(17,111)(10,0){07}{\mns}%
\multiput(37,121)(10,0){05}{\pls}%
}    \end{picture}
}
\begin{picture}(40,90)(20,0)
{\put(5,40) {$\Longrightarrow$} }\end{picture}
{    \begin{picture}(200,100)(-10,110)
\put(0,157){$\left\{\rule{0pt}{53pt} \right.$}
\put(-40,160){ {$[d/2]\!\!+\!\!1$}}
 \put(0,190) { \YoungBlock{17}{2}
}
\multiput(127,191)(10,0){04}{\pls}%
\multiput(167,191)(10,0){01}{\mns}%
\multiput(167,201)(10,0){01}{\pls}%

\put(0,140) { \YoungBlock{11}{5} }
\multiput(87,141)(10,0){03}{\mns}%
\multiput(87,151)(10,0){03}{\pls}%

\put(0,110){ \YoungBlock{8}{3} }
\multiput(07,111)(10,0){01}{\pls}%
\multiput(17,111)(10,0){07}{\mns}%
\multiput(37,121)(10,0){05}{\pls}%
    \end{picture}
}
\eee
After no \+ and \mi are left in any column,
the same can be done in any row
\bee 
{    \begin{picture}(200,100)(-10,110)
\put(0,157){$\left\{\rule{0pt}{53pt} \right.$}
\put(-40,160){ {$[d/2]\!\!+\!\!1$}}
 \put(0,190) { \YoungBlock{17}{2}
}
\multiput(117,191)(10,0){04}{\pls}%
\multiput(157,191)(10,0){02}{\mns}%

\put(0,140) { \YoungBlock{11}{5} }
\multiput(87,141)(10,0){01}{\mns}%
\multiput(97,141)(10,0){02}{\pls}%

\put(0,110){ \YoungBlock{8}{3} }
\multiput(07,111)(10,0){04}{\pls}%
\multiput(47,111)(10,0){02}{\mns}%
\multiput(67,111)(10,0){02}{\pls}%
     \end{picture}
}
\begin{picture}(40,90)(20,0)
{\put(5,40) {$\Longrightarrow$} }\end{picture}
{    \begin{picture}(200,80)(-10,110)
\put(0,157){$\left\{\rule{0pt}{53pt} \right.$}
\put(-40,160){ {$[d/2]\!\!+\!\!1$}}
 \put(0,190) { \YoungBlock{17}{2}
}
\multiput(137,191)(10,0){03}{\pls}%
\multiput(167,191)(10,0){01}{\mns}%

\put(0,140) { \YoungBlock{11}{5} }
\multiput(87,141)(10,0){01}{\mns}%
\multiput(97,141)(10,0){02}{\pls}%

\put(0,110){ \YoungBlock{8}{3} }
\multiput(07,111)(10,0){04}{\pls}%
\multiput(47,111)(10,0){02}{\mns}%
\multiput(67,111)(10,0){02}{\pls}%
    \end{picture}
} \eee
 Eventually, one is left with the relevant diagrams of the form
\bee
{    \begin{picture}(200,100)(0,110)
 \put(0,190) { \YoungBlock{17}{2}
}
\multiput(133,195)(10,0){05}{\circle*{3}}%

\put(0,140) { \YoungBlock{11}{5} }
\multiput(93,145)(10,0){03}{\circle*{3}}%

\put(0,110){ \YoungBlock{8}{3} }
\multiput(13,115)(10,0){08}{\circle*{3}}%
     \end{picture}
}
\eee
where the dots encode either only \mi\, cells or only \+ cells.
 $\blacksquare$

Lemma 8 implies in particular that the full pattern of the double
dimensional reduction of $o(d,2)$ modules into $o(d-1,1)$ modules is
not needed for the analysis of $Ker \,\cH$, \ie it is enough to use
the pattern (\ref{dimred}) of the dimensional reduction along either
$\+$ or $\mi$ extra dimension.

{}From Lemma 5 then follows that if $p<\frac{d}{2}$
($p>\frac{d}{2}$),  $Ker\, \cH$ is concentrated on the diagrams
with no $\+$ ($\mi$) cells and if
$p=\frac{d}{2}$,  $Ker\, \cH$ is concentrated  on the diagrams
with either only $\+$ or only $\mi$ cells.

Finally, we need

\noindent
\emph{\textbf{Lemma 9}\\
Let $Y_{p}\,, Y_{p-1}\subset \Omega^{rel}$
be two Lorentz Young diagrams of the class prescribed by Lemma 6 for
$p$ and $p-1$ forms, respectively, and such that $Y_{p}$ results
from $Y_{p-1}$ via addition of the form Lorentz index to the $p^{th}$
row followed by the replacement by $\mi$ of a Lorentz cell in some
 $q^{th}$ row with $q\neq p$. Then
$ \cH(Y_{p-1}) > \cH (Y_{p})$ for $q>p$ and $ \cH(Y_{p-1}) < \cH (Y_{p})$
for $q<p$.}

\bee 
{    \begin{picture}(200,110)(-20,100)
\put(0,167){$\left\{\rule{0pt}{45pt}\right.$}
\put( -5,167){${}_{p}$}
\put(70,220){$Y_{p} $}
 \put(0,190) { \YoungBlock{17}{2}}
 \multiput(167,201)(0,10){01}{\klt}
  \multiput(137,191)(0,10){01}{\klt}
\multiput(117,151)(0,10){04}{\klt}%
\multiput(87,131)(0,10){02}{\klt}%

\multiput(147,191)(10,0){03}{\mns}%

\put(0,140) { \YoungBlock{11}{5} }
\multiput(97,141)(10,0){02}{\mns}%

\put(0,120){ \YoungBlock{8}{2} }
\put(0,100){ \YoungBlock{4}{2} }
\multiput(17,101)(10,0){03}{\mns}%
\multiput(67,121)(10,0){02}{\mns}%
    \end{picture}
}
\begin{picture}(40,90)(0,0)
{\put(5,40) {$\Leftarrow\!\Rightarrow$} }\end{picture}
{    \begin{picture}(200,100)(-30,100)
\put(0,171){$\left\{\rule{0pt}{38pt}\right.$}
\put(-15,167){${}_{p-1}$}
\put(70,220){$Y_{ p-1} $}
 \put(0,190) { \YoungBlock{17}{2}}
 \multiput(167,201)(0,10){01}{\klt}
  \multiput(137,191)(0,10){01}{\klt}
\multiput(117,151)(0,10){04}{\klt}%

\multiput(147,191)(10,0){03}{\mns}%

\put(0,140) { \YoungBlock{11}{5} }
\multiput(87,141)(10,0){01}{\klt}%
\multiput(97,141)(10,0){02}{\mns}%

\put(0,120){ \YoungBlock{8}{2} }
\put(0,100){ \YoungBlock{4}{2} }
\multiput(17,101)(10,0){03}{\mns}%
\multiput(77,121)(10,0){01}{\mns}%
    \end{picture}
} \eee
Taking into account that $\Delta+p$ in Eq.~(\ref{sigma}) remains the same in the two
diagrams, the proof is analogous to that  of Lemma 8 $\blacksquare$.

{}From Lemma 9  it follows that, for $p< \frac{d}{2}$, $\cH$ may vanish only
for the maximal
possible number of $\mi$ cells below the $p^{th}$ row and no $\mi$ cells
above the $p^{th}$ row. Hence, it remains to consider the Lorentz Young
diagram
\bee \label{YD}{    \begin{picture}(160,110)(0,80)
\put(-80,140) { $ Y(L_1+1,\ldots L_p+1, L_{p+2}\ldots L_h|o(d-1,1))=$}
     \end{picture}}
{    \begin{picture}(120,140)(0,70)
\put(0,190) { \YoungBlock{12}{2} }
\multiput(117,191)(0,10){02}{\klt}
\multiput(97,141)(0,10){05}{\klt}%
\multiput(57,131)(0,10){01}{\klt}%
\put(130,200){\small{$L_1+1$}}
\put(130,190){{$L_2+1$}}
    \put(0,140) { \YoungBlock{9}{5} }
    \put(110,140){{$L_{p-1}+1$}}
    \put(70,129){{$L_{p }+1$}}
    \put(60,119){{$L_{p+2 } $}}
\put(0,130){ \YoungBlock{6}{1} }
\put(0,110){ \YoungBlock{5}{2} }
\put(0,80){ \YoungBlock{2}{3} }
\put(30,80){{$L_{h} $}}
    \end{picture}
}
 \eee
 The Lorentz Young diagram (\ref{YD}) has the
$(p+1)^{th}$ row missed compared to the original $o(d,2)$ Young
diagram because  $L_{p+1}$
$\mi$ cells are inserted below the $p^{th}$ row. This
yields
 \be \label{delt} \Delta = - L_{p+1}\,.
 \ee

Let us show that indeed $\cH=0$ in this case.
First of all we observe using (\ref{symcas}) that the contribution
of every $(p+l)^{th}$ row with $l> 0$  in (\ref{YD}) to the Lorentz
Casimir in (\ref{sigma}) cancels that of the $(p+l+1)^{th}$ row in
the $o(d,2)$ Casimir because the shift of dimension $d\to d+2$
exactly compensates the shift of the label $i$. The remaining
contributions are due to the first $p$ lines in the Lorentz Casimir,
the first $p+1$ lines of the $o(d,2)$ Casimir and  the third term in
(\ref{sigma}), \ie taking into account (\ref{delt}), \bee \cH
&=&\half( \sum_{i=1}^p [L_i (L_i +d+2 -2i ) -(L_i+1) (L_i +d+1 -2i
)]\nn\\ &+&
L_{p+1} (L_{p+1} +d -2p )-(L_{p+1}+p)(L_{p+1} +p -d))\nn\\
&=& -\sum_{i=1}^p (d+1 -2i ) -p(p -d) =0\,. \eee Thus we arrive at
the following

\noindent
{\it \textbf{Theorem}\\
Let the original $o(d,2)$--module be $F^\bL_{o(d,2)}$, and
$G^{\bL,p}_{o(d-1,1)}$ be a linear space of $o(d-1,1)$ tensors described
by the Young diagram \be \label{tens}
 Y(L_1+1,L_2+1,\ldots , L_p+1, L_{p+2},L_{p+3},\ldots |o(d-1,1))\,,
\ee
that results from $Y(\bL|o(d,2))$ by
cutting the $(p+1)^{th}$ row and adding a cell to each of the first
$p$ rows
\bee \label{coh}   
 \begin{picture}(400,150)( 20,0)

{
\begin{picture}(20,150)(0,80)
\put(0,167){$\left\{\rule{0pt}{45pt}\right.$}
\multiput(7,131)(0,10){08}{\klt}%
\put(- 10,165){$p$}
\put(23,160) { $\otimes$}
\put(153,160) { $\Rightarrow$}
\put(283,160) { $\Rightarrow$}
    \end{picture}
}

{
{    \begin{picture}(120,130)(0,80)
\put(0,190) { \YoungBlock{11}{2} }
\put(50,220){$F^\bL_{o(d,2)}$}
\put(0,140) { \YoungBlock{8}{5} }

\put(0,110){ \YoungBlock{5}{3} }
    \end{picture}
}
}

{
{    \begin{picture}(120,130)(0,80)
\put(0,190) { \YoungBlock{11}{2} }

\multiput(117,191)(0,10){02}{\klt}%

    \put(0,140) { \YoungBlock{8}{5} }
\multiput(87,141)(0,10){05}{\klt}%

\multiput(57,131)(0,10){01}{\klt}%
\put(0,110){ \YoungBlock{5}{3} }
\multiput(07,120)(10,0){05}{\krs}%
    \end{picture}
}
}

{    \begin{picture}(120,130)(0,80)
\put(50,220){$G^{\bL,p}_{o(d-1,1)}$}
\put(0,190) { \YoungBlock{12}{2} }
\multiput(117,191)(0,10){02}{\klt}
\multiput(97,141)(0,10){05}{\klt}%
\multiput(57,131)(0,10){01}{\klt}%

    \put(0,140) { \YoungBlock{9}{5} }
\put(0,130){ \YoungBlock{6}{1} }
\put(0,120){ \YoungBlock{5}{1} }
    \end{picture}
}

\end{picture}
\eee
The cohomology $H^p(\gs_-, F^\bL_{o(d,2)})$ and its conformal dimension
$\Delta$ are}\\
\bee
\label{cd1}
\ls 2p<d: \phantom{L_{p+1}=0,}\quad H^p(\gs_-, F^\bL_{o(d,2)})&=& G^{\bL,p}_{o(d-1,1)}\q
\qquad\quad\,\Delta =- L_{p+1}\,,\\
\ls 2p>d:\phantom{L_{p+1}=0,} \quad H^p(\gs_-, F^\bL_{o(d,2)})&=& G^{\bL,p}_{o(d-1,1)}\q
\qquad\quad \,\Delta = L_{p+1}\,,\\
\ls d=2p,\, L_{p+1}>0: \quad H^p(\gs_-, F^\bL_{o(d,2)})&=&
G^{\bL,p}_{o(d-1,1)} \oplus G^{\bL,p}_{o(d-1,1)}
,\,\,\, \Delta =\pm L_{p+1}\,,\\
\label{cd2}
\ls d=2p,\, L_{p+1}=0: \quad H^p(\gs_-, F^\bL_{o(d,2)})&=&
G^{\bL,p}_{o(d-1,1)} \q \qquad\quad\,\Delta =0\,. \eee

\subsection{Dynamical interpretation}
\label{dint}

According to general $\gs_-$ cohomology analysis sketched in Section
\ref{sigma-}, for a
$p$-form gauge field valued in the
$o(d,2)$--tensor module $F^\bL_{o(d,2)}$,
  $H^{q}(\gs_-,F^\bL_{o(d,2)})=G^{\bL,q}_{o(d-1,1)}$
  with $q=p-1$, $p$ and $p+1$ describe,
respectively, differential gauge symmetry parameters $\varepsilon^{dif}(x)$,
dynamical fields $\phi^{dyn}(x)$
and gauge invariant ground field strengths $C(x)$ called Weyl tensors.
Since $G^{\bL,q}_{o(d-1,1)}$ is irreducible as a Lorentz module,
all these objects have definite symmetry properties and are
traceless. In particular, the tracelessness of the dynamical fields
$\phi^{dyn}(x)$ is due to Stueckelberg symmetries
which generalize the spin two dilatation symmetry to a
general conformal field. Note that the HS dilatation symmetry
was considered in \cite{deda} for the case of spin three where it was
shown that it leaves invariant  the spin three generalized Weyl tensor
(see also \cite{Marnelius:2008er} for more examples).

Let a representative of $H^p(\gs_-,F^\bL_{o(d,2)})$
be described in
terms of the coefficients\\ $ \Omega^{a_1\ldots a_p;}{}^{A^1_1\ldots
A^1_{L_1}\,, A^2_1\ldots A^2_{L_2}\ldots}$ in Eq. (\ref{expan1}).
Let $
\Phi_{a_1\ldots a_p ;}{}_{A^1_1\ldots A^1_{L_1}\,, A^2_1\ldots
A^2_{L_2}\ldots}$ be an arbitrary auxiliary tensor, that has the
same properties as $\Omega$, \ie it belongs to $\textbf{L}$ with
respect to the indices $A^i$ and is totally antisymmetric with
respect to the form indices $a$.

For $2p\leq d$ a representative of $H^p(\gs_-,F^\bL_{o(d,2)})$ is
defined by
\bee \label{rep<} \langle\Phi| \Omega\rangle &=&
\Phi_{b_1\ldots b_p;}{}_{\,a^1_1\ldots a^1_{L_1},\ldots,
 a^p_1\ldots a^p_{L_p},{\underbrace{-\ldots -}_{L_{p+1}}},
 a^{p+2}_1\ldots a^{p+2}_{L_{p+2}}\ldots
 a^{h}_1\ldots a^{h}_{L_{h}}}\nn\\
&&\go^{a^1_1\ldots a^1_{L_1}b_1,\ldots,
 a^p_1\ldots a^p_{L_p}b_p,a^{p+2}_1\ldots a^{p+2}_{L_{p+2}},\ldots,
 a^{h}_1\ldots a^{h}_{L_{h}}}\,,
\eee
where $\go^{a^1_1\ldots a^1_{L_1+1},\ldots,
 a^p_1\ldots a^p_{L_p+1},a^{p+2}_1\ldots a^{p+2}_{L_{p+2}}\ldots
 a^{h}_1\ldots a^{h}_{L_{h}}}$
 is an arbitrary tensor valued in $G^{\bL,p}_{o(d-1,1)}$ (\ref{coh}) to
 parametrize cohomology. To strip off  $\Phi$ one should insert
 a Young projector $\Pi$ that enforces $|\Omega\rangle$
 (\ref{rep<}) to satisfy the irreducibility conditions (\ref{tr})-(\ref{length})
 of $Y(\textbf{L},o(d,2))$.

Analogously, for $2p\geq d$, a representative of
$H^p(\gs_-,F^\bL_{o(d,2)})$ is \bee \label{rep>} \langle\Phi|
\Omega\rangle  &=& \epsilon_{b_1\ldots b_{p^\prime}}
{}^{c_1\ldots c_{p}} \Phi_{c_1\ldots c_p ;\,a^1_1\ldots
a^1_{L_1},\ldots,
 a^{p^\prime}_1\ldots a^{p^\prime}_{L_{p^\prime}},{\underbrace{+\ldots +}_{L_{{p^\prime}+1}}},
 a^{{p^\prime}+2}_1\ldots a^{{p^\prime}+2}_{L_{{p^\prime}+2}},\ldots,
 a^{h}_1\ldots a^{h}_{L_{h}}}\nn\\
&&\times \go^{a^1_1\ldots a^1_{L_1}b_1,\ldots,
 a^{p^\prime}_1\ldots a^{p^\prime}_{L_{p^\prime}}b_{p^\prime},
 a^{{p^\prime}+2}_1\ldots a^{p^\prime+2}_{L_{p^\prime+2}},\ldots,
 a^{h}_1\ldots  a^{h}_{L_{h}}}\,,
\eee where $p^\prime = d-p$. For $2p=d$ the representatives
(\ref{rep<}) and (\ref{rep>}) are different at $L_{p+1} \neq 0$ and
coincide at $L_{p+1}=0$.

The gauge field transformation law can be represented in the form
\be \label{gtrp} \delta \phi^{dyn} = \cL^\bL_p\varepsilon^{dif}\,,
\ee where
 $\cL^\bL_p$ is a differential operator of order
\be
\label{qc}
q= {L_{p} - L_{p+1}+1}\,,
\ee
that maps tensor fields in
$G^{\bL,p-1}_{o(d-1,1)}$ to those in $G^{\bL,p}_{o(d-1,1)}$.
Note that in the case $L_{p} > L_{p+1}$ the operator $\cL^\bL_p$
contains more than one derivative. In this respect, the corresponding
conformal fields are analogues of the so-called partially massless
fields in $AdS_d$  originally
discovered  by Deser and Nepomechie in \cite{FirstPMWorks} and
investigated further in
\cite{Higuchi,Bengtsson:1994vn,Buchbinder:2000fy,
DeserWaldron,Zinoviev,DW,SV,skads,skads1}.
In fact,
as discussed  in Section \ref{dcs}, the system of conformal
partially massless fields considered in \cite{DW} provides a
particular example of conformal systems considered in this paper.

On the other hand,  the gauge transformation law contains one derivative in
the case where $L_p = L_{p+1}$,
\ie
the $p^{th}$ row of the original
$o(d,2)$ Young diagram is not a lowest row of some block in the
maximal block decomposition. For example this is the case in
conformal gravity  where $p=1$ and $L_1=L_2=1\,,L_3=L_4=\ldots =0$.

Analogously, in the case where the $(p+1)^{th}$ row is not a
lowest row of some block in the maximal block decomposition, the
corresponding Weyl tensor contains  one derivative of the
dynamical field $\phi^{dyn}$. In this case, the nontrivial
cohomology $H^{p+1}$ appears at the lowest level and, correspondingly,
the condition that the lowest curvature is zero analogous to
the zero-torsion condition in gravity imposes not only  constraints on
the auxiliary fields but also nontrivial differential equations on the
dynamical fields.

In fact, $\cL^\bL_p$ is determined up to an overall factor
by a chosen $o(d,2)$ Young diagram $Y(\bL|o(d,2))$ and a nonnegative
integer $p$. Its  form is determined by the Lorentz
irreducibility properties of $G^{\bL,p-1}_{o(d-1,1)}$ and
$G^{\bL,p}_{o(d-1,1)}$.  To make it explicit, it is useful to
introduce the Lorentz invariant scalar product on
$G^{\bL,p}_{o(d-1,1)}$
\be\label{lcont} (\chi\,,\phi ) = \chi^{a^1_1\ldots
a^1_{L_1+1},a^2_1\ldots a^2_{L_2+1},\ldots}\, \phi_{a^1_1\ldots
a^1_{L_1+1},a^2_1\ldots a^2_{L_2+1},\ldots} \ee
We have for $\chi\in G^{\bL,p}_{o(d-1,1)}$ and $\varepsilon\in G^{\bL,p-1}_{o(d-1,1)}$,
\bee
\label{invop} (\chi\,,\cL^\bL_p\varepsilon ) = \chi^{a^1_1\ldots
a^1_{L_1+1},a^2_1\ldots a^2_{L_2+1},\ldots, a^p_1\ldots a^p_{L_p+1},
a^{p+2}_1\ldots a^{p+2}_{L_{p+2}},
a^{p+3}_1\ldots a^{p+3}_{L_{p+3}}\ldots}\\
\times \p_{a^p_{L_{p+1}+1}}\p_{a^p_{L_{p+1}+2}}\ldots \p_{a^p_{{L_p}
\,+1}} \varepsilon_{a^1_1\ldots a^1_{L_1+1},a^2_1\ldots
a^2_{L_2+1},\ldots, a^p_1\ldots a^p_{L_{p+1}}, a^{p+2}_1\ldots
a^{p+2}_{L_{p+2}}, a^{p+3}_1\ldots a^{p+3}_{L_{p+3}}\ldots}\,.\nn
\eee
To strip off the auxiliary parameter $\chi$  one has to insert
the projector $\Pi$ to $G^{\bL,p}_{o(d-1,1)}$
\bee
\label{project}
&&(\cL^\bL_p\varepsilon)_{a^1_1\ldots
a^1_{L_1+1},a^2_1\ldots a^2_{L_2+1},\ldots, a^p_1\ldots a^p_{L_p+1},
a^{p+2}_1\ldots a^{p+2}_{L_{p+2}},a^{p+3}_1\ldots a^{p+3}_{L_{p+3}}\ldots}\\
&&=\Pi(\p_{a^p_{L_{p+1}+1}}\p_{a^p_{L_{p+1}+2}}\ldots \p_{a^p_{{L_p}
\,+1}} \varepsilon_{a^1_1\ldots a^1_{L_1+1},a^2_1\ldots
a^2_{L_2+1},\ldots, a^p_1\ldots a^p_{L_{p+1}}, a^{p+2}_1\ldots
a^{p+2}_{L_{p+2}}, a^{p+3}_1\ldots a^{p+3}_{L_{p+3}}\ldots} )\,,\nn
\eee
that removes traces and imposes Young symmetrizations. This projector
is fairly complicated in the general case. Fortunately, in many cases,
its manifest form  is not needed for practical analysis where it suffices to use
Eq.~(\ref{invop}).

The Weyl tensor
$C\in
G^{\bL,p+1}_{o(d-1,1)}$ is obtained by a similar procedure with $p$ replaced by $p+1$
\be\label{Cphi} C(\phi^{dyn})=
\cL^\bL_{p+1}\phi^{dyn} \,.
\ee
Being a part of the manifestly gauge invariant curvature (\ref{Rco}),
the  Weyl tensor is gauge invariant. This implies
\be \label{LL}
\cL^\bL_{p+1}\cL^\bL_{p}\equiv0\,. \ee
This relation with various
$p$ expresses the gauge invariance of the Weyl tensor, gauge
symmetries for gauge symmetries, Bianchi identities
\be \label{BI}
\cL^\bL_{p+2} C(\phi^{dyn})=0\,.
\ee
and syzygies
(\ie Bianchi identities for Bianchi identities).
 In fact it is not difficult to see directly
that the identity (\ref{LL}) is true because in (\ref{invop}) with
$\varepsilon = L_{p-1}^\bL \varepsilon^\prime $ some two derivatives
will be contracted with two indices of the same column in $\chi$.
 Eq.~(\ref{LL}) expresses the complex associated to $o(d,2)$-module
 homomorphisms discussed in \cite{STV}.

The space of gauge invariant field strengths
associated to the conformal Weyl tensors
coincides with the space of Weyl tensors introduced in \cite{ASV2} in the
analysis of massless mixed symmetry fields in $AdS_d$.
As discussed  in Section \ref{conclusion}, the
full $\gs_-$ cohomology in the $AdS$ case  differs from that of
the conformal case. This difference  allows
in particular second order wave equations for the fields in $AdS_d$
associated to the
additional $AdS$ cohomology. Weyl tensors in the $AdS$ theory
describe those components of the HS curvatures that
remain  nonzero on shell. Remarkably, this part of the
$AdS$ cohomology is inherited from the conformal cohomology.
For example, in
Einstein gravity, Einstein equations are associated to the
Einstein $AdS$ cohomology while the Weyl tensor contains those
components of the Riemann tensor that can be nonzero on shell.

\subsection{Examples}
\label{dcs}

\subsubsection{Maxwell theory and differential forms}
\label{forms}

$4d$ Maxwell theory  provides an example of a
gauge conformal system.
It corresponds to the case of a one-form gauge field valued
in the trivial $o(d,2)$-module  $\bL = (0,0,\ldots)$.
By Theorem of Section \ref{comp},
$G^{0,p}_{o(d-1,1)}$ (\ref{tens}) is the space of rank-$p$
antisymmetric tensors.
For the case of gauge one-form (\ie $p=1$) we have scalar gauge parameter
$\epsilon(x)$ ($G^{0,0}_{o(d-1,1)}$), vector dynamical field
$A(x)=dx^\un A_\un(x) $ $(G^{0,1}_{o(d-1,1)})$ with the gauge
transformation law
\be \delta A (x)= d \epsilon(x)
\ee
and
antisymmetric field strength $C_{a,b}(x)$ $(G^{0,2}_{o(d-1,1)})$ built
from first derivatives of $A_n(x)$
\be
\label{max} dA(x) = h^a\wedge h^b C_{a,b} (x)\,,
\ee
where $h^a$ is vierbein of $4d$ Minkowski space ($h^a=dx^a$
in the Cartesian coordinate system).
Clearly, $C_{a,b}(x)$ identifies with the Maxwell tensor.

More generally, for a $p$--form gauge field in the trivial
representation of  $o(d,2)$, we have a $(p-1)$--form  gauge parameter, a
$p$-form  dynamical field and a $(p+1)$--form ground
field strength.
Although this structure holds for any $d$,
as demonstrated in Section \ref{act}, the form of
conformal invariant field equations and, in particular, their order
depends on $d$.

\subsubsection{Conformal gravity}
\label{egr} In the case of conformal gravity, the gauge field is a
one-form valued in the adjoint $o(d,2)$--module, \ie $p=1$ and
 $\bL=(1,1,0,0,\ldots)$. According
to Theorem of Section \ref{comp}, the differential gauge
parameter  in $G^{\bL,0}_{o(d-1,1)}=
Y(1,0,0,\ldots|o(d-1,1))$  is a Lorentz vector $\varepsilon^a$,
the dynamical field in
$G^{\bL,1}_{o(d-1,1)}\sim Y(2,0,0,\ldots|o(d-1,1))$ is a
traceless symmetric tensor $g_{ab}$, and the ground field
strength in
$G^{\bL,2}_{o(d-1,1)}\sim Y(2,2,0,\ldots|o(d-1,1))$ is a Lorentz
traceless tensor $C_{ab,cd}$ described by the ``window"
Young diagram  {
\begin{picture}(14,14)
\put(0,00){\line(1,0){14}}%
\put(0,07){\line(1,0){14}}%
\put(0,14){\line(1,0){14}}%
\put(0,00){\line(0,1){14}}%
\put(7,00){\line(0,1){14}}%
\put(14,00){\line(0,1){14}}%
\end{picture}
}.
 These results agree with the standard
analysis of conformal gravity. Namely differential gauge symmetries
are linearized diffeomorphisms. The dynamical fields characterize
conformal equivalence classes of the metric, described at the
linearized level by a traceless symmetric tensor. The gauge
invariant ground field strength is the Weyl tensor.

In more familiar field--theoretical terms, the $\gs_-$ cohomological analysis is
equivalent to the following. The dilatation gauge one-form $b$
is Stueckelberg. It can be gauge fixed to zero \be \label{b0} b=0
\ee
 by a special conformal gauge transformation with the parameter
 $\tilde{\epsilon}^a (x)= \epsilon^{a-} $.
 (Here one uses that $h^a$ is nondegenerate.)
 The leftover gauge symmetries are local translations,
 Lorentz transformations and dilatations. The latter two are also Stueckelberg,
 taking away the antisymmetric and trace parts of the
 linearized fluctuation $\tilde{h}_\un{}^a$ of the vielbein.
Apart from the traceless symmetric part  of $\tilde{h}_\un{}^a$,
 the remaining fields include the Lorentz connection $\go^{ab}$ and
 special conformal connection $f^a$. These are auxiliary fields.
The Lorentz connection $\go^{ab}$ is expressed via vielbein by
 the standard zero--torsion constraint
\be \label{zerc}
R^a=0\,.
\ee
$f^a$ enters the dilatation curvature and the
Lorentz curvature. The antisymmetric part of $f_\un{}^a$ can be
adjusted to set the dilatation curvature to zero
$
R=0\,.
$
The symmetric part of $f_\un{}^a$ can be adjusted to set the trace
part of the Lorentz curvature $R_{\un\um\,;a,b}$ to zero. The part
of the Lorentz curvature $R_{\un\um\,;a,b}$  antisymmetric in three
indices is zero by virtue of the Bianchi identity that follows from
the zero--torsion constraint (\ref{zerc}). The  nonzero components
of the Lorentz curvature $R_{\un\um\,;a,b}$ are therefore contained
in its traceless part described by the window Lorentz Young diagram.
This is the Weyl tensor. All these facts can be concisely
written in the form
\be \label{Weyl} R_{ab}(x) = h^c\wedge h^d
C_{ca,db}(x)\,,
\ee
where $R_{ab}$ is the Lorentz curvature two-form and
 $C_{ab,cd}$ describes the Weyl tensor in the symmetric basis,
\ie
\be C_{ab,cd}=C_{ba,cd}=C_{ab,dc}\q
C_{(ab,c)d}=0\q C^a{}_{a,cd}=0\,.
\ee
Note that the relation with the conventional definition in the
antisymmetric basis is
\be
{C}_{[ac],[bd]}= \f{1}{3}(C_{ab,cd}-C_{cb, ad})\q
C_{ab,cd} = {C}_{[ac],[bd]}+ {C}_{[bc],[ad]}
\,.
\ee

As is well known, the condition $C_{ab,cd}=0$ implies that
metric is conformally flat, \ie the dynamical field is pure gauge.
This fact is a simple consequence of the unfolded formulation of
conformal gravity. Indeed, taking into account that, locally, any
two $o(d,2)$ flat connections are related by a $o(d,2)$ gauge
transformation and that $o(d,2)$ gauge transformations contain local
dilatations of the metric, it follows that the metric tensor is
conformally flat iff the  Weyl tensor is zero.

Let us note that the special conformal connection $f^a$ appears on
the right hand side of Einstein equations. This agrees with the
fact that (linearized) Einstein equations are not conformal
invariant because the condition $f^a=0$ or $f^a=\lambda^2  h^a$
are only Poincar\'{e} or $AdS_d$
invariant. A similar phenomenon takes place for most of conformal
systems: to obtain a unitary field-theoretical system with
second-order field equations one has to set to zero some of
components of the conformal gauge fields, that breaks down
conformal symmetry. This phenomenon is anticipated to play important
role for the study of $\gs_-$ cohomology of the Poincare' or
$AdS$ algebra in terms of conformal algebra. The
case of trivial gauge $o(d,2)$-module is exceptional
in this respect.
Hence, the case of forms considered in Section \ref{forms},
that includes $4d$ Maxwell theory, conforms with unitarity.

\subsubsection{Rectangular diagrams}
\label{rect}
Let us consider a rectangular $o(d,2)$ Young diagram of
length $s-1$ and height $h$, \ie
\be \bL = (\underbrace{{s-1,\ldots,
s-1}}_h, 0,\ldots 0\ldots 0)\,.
\ee

There are two options.
$G^{\bL,p}_{o(d-1,1)}$ with $p< h$
is the space of Lorentz tensors described by the Young diagram
 composed of two blocks. The higher one has length $s$ and height
$p$ and the lower one has length $s-1$ and height $h-p-1$
 \bee
     \begin{picture}(170,70)(0, 0)
\put(0,35){$G^{\bL,p}_{o(d-1,1)}\q p< h\q$}
     \end{picture}
 {    \begin{picture}(120,130)(0, 0)
\put(118,30){$h-p-1$}
\put(128,85){$p$}
\put(-30,45){$h-1$}
\put(115, 82.5){$\left. \rule{0pt}{18.9 pt}\right\}$}
\put(105, 37){$\left. \rule{0pt}{34.1pt}\right\}$}
\put(0, 52){$\left\{ \rule{0pt}{46.1pt}\right.$}
\put(7,10){$\underbrace{\rule{100pt}{0pt}}$}
\put(8,100){$\overbrace{\rule{108pt}{0pt}}$}
    \put(0,10) { \YoungBlock{10}{6} }
    \put(0,70) { \YoungBlock{11}{3} }
 \put(55,-10){${  s-1}$}
 \put(55,110){${  s }$}
    \end{picture}
}.
\eee
$G^{\bL,p}_{o(d-1,1)}$
with $p\geq  h$ is the space of Lorentz tensors described by the
flag Young diagram resulting from the addition the first column of height
$p$ to the original rectangular Young diagram
\bee
{\begin{picture}(150,70)(0, 0)
\put(0,35){ $G^{\bL,p}_{o(d-1,1)}\q p\geq  h\q$}\end{picture}}
{    \begin{picture}(120,100)(0, 0)
\put(130,75){$h $}
\put( -10,55){$p $}
\put(119, 85.7){$\left. \rule{0pt}{22.95 pt}\right\}$}
\put(0, 56){$\left\{ \rule{0pt}{54.1pt}\right.$}
\multiput( 7, 11)(0,10){010}{\klt}
    \put(10,70) { \YoungBlock{10}{4} }
\put(19,110){$\overbrace{\rule{98pt}{0pt}}$}
  \put(65,120){${  s }$}

    \end{picture}
}\,.
\eee
The cases with  $p=h-1$ and $p=h$ are of most interest.

In the case $p=h-1$, physical fields are described by the
rectangular Young diagram of length $s$ and height $h-1$
\bee
\label{sg}
{\begin{picture}(50,70)(0, 0)\put(0,35){ $\phi^{dyn}:$}\end{picture}}
{    \begin{picture}(120,70)(0, 0)
\put(128,35){$h-1$}
\put(115, 37){$\left. \rule{0pt}{34.1pt}\right\}$}
\put(8,10){$\underbrace{\rule{110pt}{0pt}}$}
    \put(0,10) { \YoungBlock{11}{6} }
 \put(55,-10){${  s}$}

    \end{picture}
}.
\eee
Weyl tensor is described by the
rectangular Young diagram of length $s$ and height $h$
\bee
\label{sw}
{\begin{picture}(30,70)(0, 0)\put(0,35){ $C:$}\end{picture}}
 {    \begin{picture}(120,70)(0, 0)
\put(128,35){$h$}
\put(115, 37){$\left. \rule{0pt}{34.1pt}\right\}$}
\put(8,10){$\underbrace{\rule{110pt}{0pt}}$}
    \put(0,10) { \YoungBlock{11}{6} }
 \put(55,-10){${  s}$}

    \end{picture}
}
\eee
and gauge parameter $\varepsilon^{dif}$ is described by the
diagram
 \bee
     \begin{picture}( 40,70)(0, 0)
\put(0,35){$ \varepsilon^{dif}  :  $}
     \end{picture}
 {    \begin{picture}(130,90)(0, 0)
\put(128,40){$h-2$}
\put(117, 47){$\left. \rule{0pt}{34.1pt}\right\}$}
\put(7,10){$\underbrace{\rule{100pt}{0pt}}$}
\put(8,80){$\overbrace{\rule{108pt}{0pt}}$}
    \put(0,10) { \YoungBlock{10}{1} }
    \put(0,20) { \YoungBlock{11}{6} }
 \put(55,-10){${  s-1}$}
 \put(55,90){${  s }$}
    \end{picture}
}.
\eee
These fields generalize the
single row symmetric conformal HS fields of \cite{FT,segal},
which correspond to the case of $h=2$, $p=1$ as well as the
case of $h=\frac{d}{2},$ $p=h-1$ considered recently in
\cite{Marnelius:2009uw}, to an arbitrary rectangular block.
The example of forms of
Subsection \ref{forms} corresponds to $s=1$. $4d$ conformal
gravity has $h=s=2$, $p=1$. In all cases with $p=h-1$,
the Weyl tensor $C$ contains $s$ derivatives of $\phi^{dyn}$
and the gauge transformation law of $\phi^{dyn}$ contains one
derivative of the gauge parameter $\varepsilon^{dif}$.

In the case $p=h$, physical fields are described by the
rectangular Young diagram of length $s$ and height $h$
\bee
\label{fbl}
{\begin{picture}(50,70)(0, 0)\put(0,35){ $\phi^{dyn}:$}\end{picture}}
{    \begin{picture}(120,70)(0, 0)
\put(128,35){$h$}
\put(115, 37){$\left. \rule{0pt}{34.1pt}\right\}$}
\put(8,10){$\underbrace{\rule{110pt}{0pt}}$}
    \put(0,10) { \YoungBlock{11}{6} }
 \put(55,-10){${  s}$}

    \end{picture}
}.
\eee
Weyl tensor is described by the diagram
\bee
\label{cbl}
{\begin{picture}(30,70)(20, 0)\put(0,35){ $C:$}\end{picture}}
{    \begin{picture}(120,90)(0, 0)
\put(-15,35){$h$}
\put(0,37){$\left\{\rule{0pt}{34.1pt}\right.$}
\put(8,70){$\overbrace{\rule{110pt}{0pt}}$}
    \put(0,10) { \YoungBlock{11}{6} }
    \put(0,0) { \YoungBlock{1}{1} }
 \put(55,80){${  s}$}
    \end{picture}
}.
\eee
 The gauge parameter is described by the rectangular Young diagram of
length $s$ and height $h-1$. The Weyl tensor $C$ contains one
derivative of the dynamical field $\phi^{dyn}$, while the gauge
transformation of the latter contains $s$ derivatives of
$\varepsilon^{dif}$. This
example generalizes to any $h$ partially massless conformal
fields considered in \cite{DW,SV}, which correspond to the case of
$h=1$. Again, the degenerate case of $s=1$ describes differential
forms.

Finally,  consider the case of a one-form gauge field
 in the single column representation of height $h$.
 In this case, dynamical field
 is the hook of height $h-1$
\bee
{\begin{picture}(50,70)(0, 0)\put(0,35){ $\phi^{dyn}:$}\end{picture}}
{    \begin{picture}(120,60)(0, 0)
\put(40,35){$h-1$}
\put(29, 32){$\left. \rule{0pt}{30.1pt}\right\}$}
    \put(0,10) { \YoungBlock{1}{5} }
    \put(10,50) { \YoungBlock{1}{1} }
    \end{picture}
}
\eee
the gauge parameter is a rank $h-1$ antisymmetric tensor
\bee
{\begin{picture}(50,70)(0, 0)\put(0,35){ $\varepsilon^{dif}:$}\end{picture}}
{    \begin{picture}(120,60)(0, 0)
\put(30,35){$h-1$}
\put(19, 32){$\left. \rule{0pt}{30.1pt}\right\}$}
     \put(0,10) { \YoungBlock{1}{5} }
    \end{picture}
}
\eee
and the
gauge invariant field strength is a two column traceless tensor with
$h-1$ cells in the first column and two cells in the second column
\bee
{\begin{picture}(40,70)(0, 0)\put(0,35){ $C:$}\end{picture}}
{    \begin{picture}(120,60)(0, 0)
\put(48,35){$h-1$}
\put(35, 32){$\left. \rule{0pt}{30.1pt}\right\}$}
    \put(0,10) { \YoungBlock{1}{5} }
    \put(10,40) { \YoungBlock{1}{2} }
    \end{picture}
}.
\eee

 For the particular case of
$h=3$ this gives three-cell hook dynamical field with antisymmetric
tensor
 \begin{picture}(7,14)
\put(0,00){\line(1,0){7}}%
\put(0,07){\line(1,0){7}}%
\put(0,14){\line(1,0){7}}%
\put(0,00){\line(0,1){14}}%
\put(7,00){\line(0,1){14}}%
\end{picture}
 as a gauge parameter and window
 \begin{picture}(14,14)
\put(0,00){\line(1,0){14}}%
\put(0,07){\line(1,0){14}}%
\put(0,14){\line(1,0){14}}%
\put(0,00){\line(0,1){14}}%
\put(7,00){\line(0,1){14}}%
\put(14,00){\line(0,1){14}}%
\end{picture}
  as the generalized
Weyl tensor. Both the gauge transformation law and the expression
for the Weyl tensor in terms of the dynamical fields contain
one derivative. Let us note that this pattern is
analogous to that of the hook field in flat
\cite{Curtright:1980yk,Zinoviev:2002ye,Zinoviev:2003dd,Zinoviev:2008ve}
and $AdS_d$ \cite{BMV} spaces.
The difference is that in the conformal case the dynamical field is
traceless while in the $AdS$ case it is not \cite{BMV}. This is
 analogous to the case of gravity where the dynamical field is
traceless in the conformal case and traceful in the $AdS$ or
Poincare' case. On the
other hand, in the Poincar\'{e} invariant case
 the hook system acquires an additional gauge symmetry
with the symmetric tensor gauge parameter
\cite{Curtright:1980yk,BMV}.

\subsection{Structure of unfolded conformal equations}
\label{stue}
{}From the analysis of $\sigma_-$
cohomology it follows that, in the
system described by a gauge $p$-form field with
$2p\leq d$, all components of the curvature
$p+1$-forms with conformal dimensions
$ \Delta< \Delta(H^{p+1}(\gs_-, F^\bL_{o(d,2)})) $ can
be set to zero  imposing no restrictions on the
dynamical fields, \ie these equations are zero-torsion-like
constraints that express auxiliary fields via derivatives of
the dynamical fields. Let us write these equations symbolically  as
\be \label{R-} R^{\ldots}_-
=0\q \Delta(R_-)<\Delta(H^{p+1}(\gs_-, F^\bL_{o(d,2)}))\,.
\ee
The lowest conformal dimension of curvatures components $R_{C}$ that
cannot be set to zero by some choice of constraints on the
connections is $\Delta(H^{p+1}(\gs_-, F^\bL_{o(d,2)})). $
Then the equation
\be
\label{Req}
R^{\cdots}_{C}=\underbrace{h^{\cdots}\wedge\ldots h^{\cdots}}_{p+1}
C^{\cdots}
\ee
is a constraint that expresses some auxiliary fields
contained in the $p$-form gauge field $W \in \cF^{\bL,p}_{o(d,2)}$
and the generalized Weyl zero-form $C\in H^{p+1}(\gs_-,
F^\bL_{o(d,2)})$ via derivatives  of the dynamical fields
$\phi^{dyn}\in H^{p}(\gs_-, F^\bL_{o(d,2)})$ imposing no
restrictions on the latter.

More precisely, the equation (\ref{Req}) reads as \bee \label{CEQ}
\Phi_{A^1_1\ldots A^1_{L_1},\ldots,
 A^{h}_1\ldots A^{h}_{L_{h}}}&&\ls
 R_{C}^{A^1_1\ldots A^1_{L_1},\ldots,
 A^{h}_1\ldots A^{h}_{L_{h}}}=
\Phi_{a^1_1\ldots a^1_{L_1},\ldots,
 a^{p+1}_1\ldots a^{p+1}_{L_{p+1}},{\underbrace{-\ldots -}_{L_{p+2}}},
 a^{p+3}_1\ldots a^{p+3}_{L_{p+3}},\ldots,
 a^{h}_1\ldots a^{h}_{L_{h}}}\nn\\
&&h_{b_1}\wedge \ldots \wedge h_{b_{p+1}}C^{a^1_1\ldots
a^1_{L_1}b_1,\ldots,
 a^{p+1}_1\ldots a^{p+1}_{L_p}b_{p+1},a^{p+3}_1\ldots a^{p+3}_{L_{p+2}},\ldots,
 a^{h}_1\ldots a^{h}_{L_{h}}}\,,
\eee where $\Phi_{A^1_1\ldots A^1_{L_1},\ldots,
 A^{h}_1\ldots A^{h}_{L_{h}}} \in F^\bL_{o(d,2)}$ is an auxiliary parameter
 introduced to avoid complicated projectors. From this formula it follows that
nonzero components of the $p+1$--form curvature
$R_{C}^{A^1_1\ldots A^1_{L_1},\ldots,A^{h}_1\ldots A^{h}_{L_{h}}}$ belong to
the space $U^{\bL,p}_{o(d-1,1)}$ of fiber Lorentz tensors
described by the Young diagram that
results from the $o(d,2)$ Young diagram $F^\bL_{o(d,2)}$ by cutting
its $(p+2)^{th}$ row
\bee \label{U}   %
 \begin{picture}(400,150)( 10,0)

{
\begin{picture}(20,150)(0,80)
\put(0,177){$\left\{\rule{0pt}{32pt}\right.$}
\put( -10,185){$p$}
\put(170,167){$\left\{\rule{0pt}{45pt}\right.$}
\put(  140,175){$p+2$}

  \put(113,160) {$\Rightarrow$}
  \put(283,160) { $\Rightarrow$}
    \end{picture}
}

{ {    \begin{picture}(120,130)(30,80) \put(0,190) {
\YoungBlock{11}{2} } \put(50,220){$F^\bL_{o(d,2)}$}

\put(0,140) { \YoungBlock{8}{5} }

\put(0,120){ \YoungBlock{5}{2} }

\put(0,110){ \YoungBlock{3}{1} }
    \end{picture}
} }

{ {    \begin{picture}(120,130)(0,80) \put(0,190) {
\YoungBlock{11}{2} }

    \put(0,140) { \YoungBlock{8}{5} }

\put(0,120){ \YoungBlock{5}{2} }
\put(0,110){ \YoungBlock{3}{1} }
\multiput(07,130)(10,0){05}{\krs}%
    \end{picture}
} }

{    \begin{picture}(120,130)(0,80)
\put(50,220){$U^{\bL,p}_{o(d-1,1)}$} \put(0,190) {
\YoungBlock{11}{2} }

    \put(0,140) { \YoungBlock{8}{5} }
 \put(0,120){ \YoungBlock{3}{1} }
\put(0,130){ \YoungBlock{5}{1} }

    \end{picture}
}

\end{picture}
\eee (Let us stress that this prescription refers to the fiber
indices, while the differential form indices remain untouched.)

A useful viewpoint is to interpret the Maxwell tensor $C_{a,b}(x)$
in (\ref{max}),  Weyl tensor $C_{ab,cd}(x)$ in
(\ref{Weyl}) and generalized Weyl tensors $C$  in (\ref{Req}) as new
zero-form fields valued in $H^{p+1}(\gs_-,F^\bL_{o(d,2)})$.  With this
interpretation one can say that the generalized Weyl tensor fills
in nontrivial $\sigma_-$ cohomology in the gauge curvature so that the equation
(\ref{Req}) becomes an algebraic constraint that expresses generalized
Weyl tensors and, may be, some other auxiliary gauge fields  via
derivatives of the dynamical fields. Being defined in terms of the gauge invariant
curvatures, the generalized Weyl tensors are manifestly gauge invariant.

In fact, the generalized Weyl tensor $C$ is the ground (primary) field of an
infinite dimensional $o(d,2)$-module called Weyl module. To uncover
its structure one has to study  Bianchi identities that by virtue of
the definition (\ref{Req}) of the generalized Weyl tensor
impose restrictions on the derivatives of
$C(x)$. Denoting the unrestricted combinations of derivatives of $C(x)$
as new fields $\cC^I(x)$,  eventually  leads to the equations of the form
\be
\label{Weq} \D_0 \cC^I(x) =0\,,
\ee
where the zero--forms $\cC^I(x)$
take values in some infinite dimensional $o(d,2)$--module $W^{off}$ (index
$I$), which we call {\it off-shell Weyl module}, while the
generalized Weyl tensor $C(x)$ is valued in its vacuum subspace
 with minimal conformal dimension. $\D_0$ is the $o(d,2)$ covariant
derivative in $W^{off}$, built from the vacuum $o(d,2)$
connection $W_0$. The higher components of $\cC^I$, \ie those that
have higher conformal dimension than the ground Weyl tensor, are
expressed by virtue of (\ref{Weq}) via derivatives of the latter
(see also Section \ref{weylmod}).

In terms of zero-forms $\cC^I$,
the full unfolded system consists of the
covariant constancy equation (\ref{Weq})  along with the equations
that extend Eqs.~(\ref{R-}) and (\ref{Req}) to all
conformal dimensions
\be \label{runf}
R^{\cdots}=\underbrace{W_0^{\cdots}\wedge\ldots W_0^{\cdots}}_{p+1}
\cC^{\cdots}\,. \ee

The equations  $(\ref{Weq})$ and (\ref{runf}) are formally
consistent for any $o(d,2)$ flat connection and provide a particular
example of the linearized unfolded system (\ref{Rco}), (\ref{freq}).
This system is off-shell because it imposes no differential
conditions on the physical field $\phi^{dyn}$. Nontrivial conformal
equations result from the farther reduction of $W^{off}$ as discussed
in Section \ref{weylmod}.

Conformal dimension of the Weyl tensor valued in the Lorentz module
$G^{\bL,p+1}_{o(d-1,1)}$ can be read off the equation (\ref{Req})
\bee
\label{wdim}
\Delta C &=& -L_{p+2} +{p+1}\q 2(p+1)\leq d\,,\nn\\
\Delta C &=& \quad L_{p+2} +{p+1}\q 2(p+1)\geq d\,.
\eee
Here the terms
with $L_{p+2}$ result from $L_{p+2}$ \mi or \+ cells in
 the original $o(d,2)$ tensor module $\bL$
where the $p$-form connection $W_1$ and the $p+1$-form curvatures
$R_1$ are valued. The $p$-dependent terms come from the
contribution of the background frame one-forms in (\ref{Req}), since
being associated to translations, each of the vielbein
 one-forms carries conformal dimension 1. In the case of $2(p+1)=d$ the Weyl
tensors with both conformal dimensions are present.

Analogously, representing the gauge $p$-form $W_1$ in terms of the
zero-form dynamical field $\phi^{dyn}$ and the $(p-1)$-form gauge
parameter in terms of the zero-form differential gauge parameter
$\epsilon^{dif}$ as
\be
W_1^{\cdots}=\underbrace{h^{\cdots}\wedge\ldots h^{\cdots}}_{p}
\phi^{dyn}{}^{\cdots}\q
\epsilon_1^{\cdots}=\underbrace{h^{\cdots}\wedge\ldots
h^{\cdots}}_{p-1} \epsilon^{dif}{}^{\cdots}\,, \ee
implies that
\bee
\label{W1eq}
\Delta \phi^{dyn} &=& -L_{p+1} +{p}\q 2p\leq d\,,\nn\\
\Delta \phi^{dyn} &=& \quad L_{p+1} +{p}\q 2p\geq d\,. \eee \bee
\Delta \epsilon^{dif} &=& -L_{p} +{p-1}\q 2(p-1)\leq d\,,\nn\\
\Delta \epsilon^{dif} &=& \quad L_{p} +{p-1}\q 2(p-1)\geq d\,. \eee

 As explained in Section \ref{civ}, that the system of equations
  (\ref{Weq}) and (\ref{runf}) along with (\ref{r0}) is unfolded
implies that it is conformal invariant. By the general
formula (\ref{gltr}), the transformation law consists of two parts.
The standard part, that results from the terms linear in $W_0$, is
 the transformation law in the module where $W_1$ is
valued. Additional terms, that mix forms of different degrees,
originate from the terms nonlinear in $W_0$. This
modification leads in particular to the nontrivial conformal
transformation for the spin one field, described by a one-form
valued in the trivial $o(d,2)$-module. Indeed,
from the unfolded equation (\ref{max}) it follows that
\be
\delta A_\un = 2 \varepsilon^n(x) h_\un^m C_{nm}(x)=
2 \Big (\varepsilon^m(x) \partial_m A_\un
+\partial_\un ( \varepsilon^m(x)) A_m
- \partial_\un ( \varepsilon^m(x) A_m) \Big )\,,
\ee
where the last terms describes a field-dependent gauge transformation.

An important consequence of Eq. (\ref{Weq}) and general formula
(\ref{gltr}) applied to zero-forms
\be \delta
\cC^I= (\epsilon (x)\cdot \cC)^I\,
\ee
is that generalized Weyl
tensors are invariant under special conformal gauge transformations, \ie
generalized Weyl tensors are primary conformal fields.
Indeed, since Weyl tensor
has minimal conformal dimension $\Delta$ in the Weyl module, its
special conformal transformation is zero
because it increases conformal dimension.
Let us stress that, being a consequence of the unfolded
formulation,  the proof of this fundamental property does not
use the manifest form of $C(\phi^{dyn})$.

\section{Conformal actions}
\label{act}

\subsection{Generalities}
\label{genpr}

Manifestly gauge invariant actions can be easily
constructed from gauge invariant tensors $C(\phi^{dyn})$
and their derivatives $E(C)$
\be
S=\int d^d x L(E(C(\phi^{dyn})))\,.
\ee
However, most of these actions are not conformal.
Let us consider some examples.

The action of four dimensional Maxwell electrodynamics is
\be
S=\half\int d^4 x\, C^{a,b}(\phi^{dyn})C_{a,b}(\phi^{dyn})\,,
\ee
where $C_{a,b}(\phi^{dyn})$ is the
Maxwell tensor, which in our terms is the generalized Weyl
tensor for the one-form conformal field $A_\un=\phi_\un^{dyn}$ in the trivial
representation of the conformal group.

Conformal action of linearized $4d$ conformal gravity is
\be S=\half \int d^4 x\,
C^{ab,cd}(\phi^{dyn})C_{ab,cd}(\phi^{dyn})\,,
\ee
where $\phi_{ab}^{dyn}$ is the traceless part of  linearized metric and
$C_{ab,cd}(\phi^{dyn})$ is the linearized Weyl tensor.

These actions are both gauge invariant and conformal
invariant, as well as the actions of $4d$ conformal
symmetric HS fields \cite{FT}, which in our notations are
\be
\label{achs}
S=\half \int d^4 x\,
C^{a_1\ldots a_s,b_1 \ldots b_s}(\phi^{dyn})
C_{a_1\ldots a_s,b_1 \ldots b_s}(\phi^{dyn})\,.
\ee
$\phi^{dyn}_{a_1\ldots a_s}$ are metric-like
HS fields that are  rank $s$ traceless symmetric tensors
where $s$ is spin.

However,   Maxwell action and HS actions (\ref{achs})
are not conformal  beyond $d=4$ \cite{segal} because
the dilatation invariance requires conformal dimension $\Delta_L $
of a Lagrangian to match that of the volume element.
Namely, as we demonstrate shortly, the necessary
condition is
$
 \Delta_L = d\,.
$
For  a Lagrangian
\be \label{LR} L= \Psi_1 E(\p)\Psi_2
\ee
bilinear in the fields $\Psi_1$
and $\Psi_2$ of conformal dimensions $\Delta_1$ and
$\Delta_2$, respectively,
$ \Delta_L = \Delta_1 +\Delta_2 +
q
$,
provided that the differential operator $E(\p)$ has order $q$.
Hence the dilatation invariance requires
\be\label{pdd} q= d
-\Delta_1 -\Delta_2\,.
\ee
Since conformal dimensions (\ref{cd1})-(\ref{cd2}) of generalized
 Weyl tensors are $d$--independent, this
implies that $q$ should increase with $d$. The condition (\ref{pdd})
is necessary but not sufficient, however, because it does not
guarantee invariance of the Lagrangian under special conformal
transformations. As demonstrated in Section \ref{sct}, the latter
condition restricts strongly
$E(\p)$ and  $\Delta_{1,2}$.

The simplest way to check global conformal invariance
is to use that global conformal symmetry transformations
in Minkowski geometry  are particular $o(d,2)$ gauge
transformations (\ref{glsec})-(\ref{gltr}). A nice feature of
this approach is that, by their definition, the global transformations
leave invariant background (vacuum) connections in the
Lagrangian. Neither, they act on the space-time coordinates $x^\un$.
As a result, everything is determined by the conformal properties of
the fields that enter Lagrangian.

Let us analyze the transformation of a Lagrangian under the global
transformations (\ref{glsec})-(\ref{gltr}). We shall
discard  total derivatives, hence checking invariance of the action.
(Note that the total derivative terms are artifacts of our formalism in
which space-time coordinates are not affected by the global symmetry
transformations.) In this analysis one should  take into account
that, in accordance with the form of unfolded equations $(d +dx^a P_a)C=0$
in Cartesian coordinates with the vielbein one-form $h^a=dx^a$,
the translation generator  is
\be
P_a =- \frac{\p}{\p x^a}\,.
\ee

Invariance under global translations requires $E$ in (\ref{LR}) be
$x$-independent. The variation under global
dilatation with the constant parameter $\epsilon$ gives, taking into
account (\ref{gldil}) and (\ref{gltr}),
\be \delta L =
\epsilon(\Delta_1+\Delta_2+q - d)L+
 \epsilon \p_n (x^n L)\,.
\ee
Hence, the condition (\ref{pdd}) indeed guarantees global dilatation
invariance.

Lorentz invariance requires Lorentz indices be
contracted by the Lorentz invariant metric or epsilon symbol. In this paper we mostly
consider $p$-even models free of the epsilon symbol.

Let  Lorentz tensor fields
$\Psi(x)$ be realized as Fock vectors analogous to (\ref{expan})
\be
\label{expanl} |\Psi(x)\rangle = \sum_{l_1\geq 0,l_2\geq 0,\ldots
}\frac{1}{\sqrt{l_1! l_2 !\ldots}} \Psi^{a^1_1\ldots a^1_{l_1}\,,
a^2_1\ldots a^2_{l_2}\ldots}(x) a^{\dagger{}1}_{a^1_1}\ldots
a^{\dagger{}1}_{a^1_{l_1} } a^{\dagger{}2}_{a^2_1}\ldots
a^{\dagger{}2}_{a^2_{l_2}}\ldots|0\rangle\,.
\ee
Lorentz generators are given by the formula analogous to
(\ref{t2})
\be
\label{tl2} T^{ab} =\sum_i \left (a^{\dagger{}a i} a^b_i -
a^{\dagger{}bi} a^a_i \right )\,.
\ee
Also the field $|\Psi\rangle$ and
its conjugate $\langle\Psi |$ are assumed to be Lorentz irreducible, satisfying the
conditions
\be
\label{trlor} t_{ij} |\Psi\rangle =0\,,
\ee
\be \label{asymlor}
t^i{}_j |\Psi\rangle =0 \qquad j>i\,, \ee \be \label{lengthlor}
(t^i{}_i-l_i ) |\Psi\rangle= 0 \q \mbox{no summation over $i$,}
\ee
where
\be \label{t} t^{ij} = a^{\dagger{}ai} a^{\dagger{}b
j}\eta_{ab}\q t^i{}_j = \half \{ a^{\dagger{}a i}\,,a^b_j\}
\eta_{ab}\q t_{ij} = a^a_i a^b_j\eta_{ab} \ee
and
non-negative integers $l_i$ satisfy $l_j \geq  l_i$ for $i>j$,
characterizing the lengths of rows of a Young tableau associated to
the  Lorentz module $V^{\bf \,l}$ (${\bf l}=(l_1,l_2,\ldots l_h )$),
where $|\Psi\rangle$ is valued.
Analogously,
\be\langle\Psi| t^{ij} =0\q
\langle\Psi| t^i{}_j  =0 \qquad j<i\q \label{lengthlorc}
\langle\Psi| (t^i{}_i - l_i )=0\,.
\ee

In these terms, a general
Poincar\'{e} invariant Lagrangian for $|\Psi_{1}\rangle $
and $|\Psi_{2}\rangle $ valued in
the same Lorentz module $V$ reads
\be
\label{lagr} L= \langle \Psi_1 | E | \Psi_2 \rangle\,, \ee where \be
\label{RD} E=E(\square , \Theta^i{}_j ) \ee
and
 \be \label{DU}
 \square =\p^a \p_a\q
\Theta^i{}_j = a^{\dagger{}i\,a} a_j^b\p_a\p_b\,.
\ee

Let us note that the operators $\Theta^i{}_j$ satisfy ``almost
$gl_h$" commutation relations
\be [\Theta^i{}_j \,, \Theta^l{}_k ]=
\square \left (\delta_j^l \Theta^i{}_k - \delta_k^i \Theta^l{}_j
\right )\q [\Theta^i{}_j \,, \square]=0\,.
\ee
These relations
extend to ``almost $sp(2h)$" by adding the operators
\be \label{Dpm}
\Theta^{ij} = a^{\dagger{}i\,a} a^{\dagger{}j\,b}\p_a\p_b\q
\Theta_{ij} = a_{i}^a a_j^b \p_a\p_b\,,
\ee
that satisfy \be
[\Theta_{ij}\,,\Theta^{nm}]=\square( \delta_i^n \Theta^m{}_j+
\delta_j^n \Theta^m{}_i+\delta_i^m \Theta^n{}_j+\delta_j^m
\Theta^n{}_i )
\ee
(other commutators are zero). A nonlocal
rescaling to $sp(2h)$  is $(\Theta^i{}_j,
\Theta^{ij},\Theta_{ij})\longrightarrow \square^{-1}(\theta^i{}_j,
\theta^{ij},\theta_{ij})$.

It is easy to see that any operator of the form (\ref{RD}) has
the following commutation relations with the Lorentz generator
(\ref{tl2})
\be \label{comlor} [T_{ab}\,, E]=E_a \p_b - E_b \p_a\q
 E_a = [E\,,x_a ]=\f{\p E(\p)}{\p \p^a}\,.
\ee
The meaning of
(\ref{comlor}) is that, as is obvious from (\ref{RD}) and (\ref{DU}), Lorentz rotation
of  indices carried by
oscillators is equivalent to that of indices carried by derivatives. Another useful
relation is
\be
\label{degr} E^a \p_a = q E\,,
\ee
where $q$ is the order of the differential operator $E(\p)$. Note that in the case where
$\Psi_1, \Psi_2\in V$  and the Lagrangian is free
of the epsilon symbol, $q$ is even because it equals to $2dim\, V$ minus twice a
number of metric tensors that contract Lorentz indices.

Using these formulae along with (\ref{gllor}) and
(\ref{gltr}) it is easy to see that, as anticipated, any Lagrangian
(\ref{lagr}) is Lorentz invariant up to  a total derivative, hence leading
to the Poincar\'{e} invariant action.
Eq.~(\ref{pdd}) guarantees scaling invariance.
Analysis of special conformal symmetry is less trivial.
Note that, from the action perspective, one difference between special conformal
transformations and translations is that the actions are invariant
under translations because  the integration  allows us to discard
 total derivatives $\delta L\sim P_a \Sigma^a$. In the case
of special conformal transformations, the terms $\delta L\sim K^a
\Sigma_a$ cannot be integrated out.

\subsection{Special conformal symmetry}
\label{sct}
Now we are in a position to derive the full conformal invariance conditions on $E$
at the condition that the
fields $\Psi_{1,2}$ in the Lagrangian (\ref{lagr}) are primaries,
\ie
that they do not transform directly under special
conformal gauge transformations. The global special
conformal transformation
of primary  fields is entirely described by the $x$-dependent
dilatations (\ref{gldil}), Lorentz transformations (\ref{gllor}) and
translations (\ref{gltr}) with the parameter $\tilde{\varepsilon}^a$.
It should be stressed that the condition that
$\Psi_{1,2}$  are primaries does not restrict the class
of conformal Lagrangians because, as is customary in conformal field theory,
all other fields are descendants (\ie derivatives)
of the primary fields. A contribution of descendants
can  still be described by the Lagrangian (\ref{lagr}) with an appropriate
differential operator $E$.

In the unfolded formulation of conformal theory, primary fields are valued
in $H(\sigma_-)$ provided that $\sigma_-$ is defined so that it does not mix
differential forms of different degrees. (This is the case where the terms
polylinear in $W_0$ in (\ref{Rco}) are not included in the definition of
$\sigma_-$ so that it has the form (\ref{s-}) in the respective module of
the conformal algebra.) As explained in Subsection (\ref{sigma-}), all
other fields are descendants that are expressed via derivatives of the
primaries by the unfolded equations. In the case of conformal gauge fields
studied in Section \ref{Sigma}, there are two types of primary fields for every
conformal system, namely, dynamical fields $\phi^{dyn}$ and Weyl tensors
$C(\phi^{dyn})$.

An elementary computation shows that, up to total derivatives,
the special conformal variation of the Lagrangian
(\ref{lagr}) with primaries $\Psi_{1,2}$ is
\be
\label{confvar}
\delta
L = \tilde{\varepsilon}^a \langle \Psi_1 | \left( -(q-1+\Delta_2)E_a
+\half E_b{}^b \p_a + E^b M_{ba} \right ) | \Psi_2 \rangle\q
E_b{}^b = \frac{\p^2 E}{\p\, \p^b \p\,\p_b}\,,
\ee
where we made use of the dilatation invariance condition
(\ref{pdd}). Thus, any $E$, that satisfies (\ref{pdd}) and
\be
\label{spec} -(q-1+\Delta_2)E_a +\half E_b{}^b \p_a + E^b M_{ba}\sim 0\,,
\ee
where $\sim$ implies equivalence  up to terms that are zero
by virtue of the irreducibility conditions (\ref{trlor})-(\ref{lengthlorc}),
gives rise to a conformal invariant Lagrangian (\ref{lagr}).
This equation has interesting properties summarized in Appendix,
though they are not used directly in this paper.

Let us look for a solution of Eq.~(\ref{spec}) of the form
\be E=
E(\square\,, \Theta_i )\q\quad \Theta_i = \Theta^i{}_i\q
\mbox{no summation over repeated indices.}
\ee
 It is not difficult to see
that this form of $E$ can always be achieved for
Lorentz fields that satisfy the irreducibility conditions
(\ref{trlor})-(\ref{lengthlorc}).

Let  $E$ be represented in the form
\be\label{R} E=
 \oint dt\,  ds\, \rho(s,t) {\Large :}
 \exp t(\square +\sum_{i=1}^h s^i \Theta_i)
{\Large :}\,, \ee
where  normal ordering is such that $a^{\dagger{}ai}$  and
$a^a_i$ are moved to the left and right, respectively. The
integration is defined via Cauchy formula normalized in such
a way that
\be\label{oint} \oint dw w^{-1} = 1\q \oint d w
w^n =0 \,,\quad n\neq -1\,
\ee
for $w=t$ or $w=s^i$.
The following useful identities result from partial integration
\be \oint dt ds \sigma(s,t)(\square +\sum_{i=1}^h s^i \Theta_i)
\exp t(\square +\sum_{i=1}^h s^i \Theta_i) =
- \oint dt ds \f{\p}{\p t} \sigma(s,t)\exp t(\square +\sum_{i=1}^h
s^i \Theta_i) \,,
\ee
\be \oint dt\,
ds\,\sigma(s,t)\, t \Theta_j \exp  t(\square +\sum_{i=1}^h s^i \Theta_i )
= -
\oint dt ds \f{\p}{\p s^j} \sigma(s,t) \exp t(\square +\sum_{i=1}^h s^i \Theta_i)
\,. \ee

The computation of (\ref{confvar}) is not too hard. It is most conveniently
accomplished  in terms in the star-product formalism with the normal-ordered
star-product $*$ induced by the normal ordering introduced above.
In this computation one sets to zero terms of the form $A*t^r$ or
$t^l *A$ where $t^r$ and $t^l$ are operators that are zero by virtue
of (\ref{trlor})-(\ref{lengthlor}) or (\ref{lengthlorc}),
respectively. The final result is
\bee
\label{specrez} &&\ls\langle \Psi_2 |\half E_b{}^b \p_a + E^b
M_{ba}-(q-1+\Delta_2)E_a |\Psi_1 \rangle = \oint dt ds \, t
  \langle \Psi_2 |: \exp(t (\square +\sum_{i=1}^h s^i \Theta_i))
\nn\\
&&\ls\sum_i \Big \{ \Big (d -2t\f{\p}{\p t} -2(\Delta_2+q+1)\Big )
[\p_a +\half s^i (a_a^{+i}a_i^b  + a^{+bi}a_{a\,i} )\p_b]\\
&&\ls+ \Big(  s^i(l_i+\f{d}{2}+h-2i  -t\f{\p}{\p t}
 +\sum_{j\geq i} s^j\f{\p}{\p s^j})
+2 + \sum_{j< i}s^j(1+s^j \f{\p}{\p s^j}) \Big )(a^{+bi}a_{a\,i} -
a_a^{+i}a_i^b) \p_b
\Big \}:\nn\\
&&
\qquad \sigma(s,t) |\Psi_1 \rangle \nn .
\eee

The condition that the first term on the right hand side of
(\ref{specrez}) is zero demands
\be
\label{sigmast}
\sigma(s,t)=
t^{(\f{d}{2}-\Delta_2- q -1)}\sigma(s)\,.
\ee
Since a power of $t$ in
(\ref{R}) determines by virtue of (\ref{oint}) an order of the
differential operator in the Lagrangian, this in turn implies that
$ q=2\Delta_2 +2q -d\,.
$
Comparing this with the scaling
invariance condition (\ref{pdd}), we  obtain that
 $\Delta_1=\Delta_2$. Hence, in the sequel we set
\be \label{confc}
\Delta_1=\Delta_2=\Delta = \half (d-q)\,. \ee
Since
$\Delta_1=\Delta_2$, one can choose $\Psi_1 = \Psi_2$.
Because ${q}$ is even, the condition (\ref{confc})
implies that $\Psi_{1,2}$ has integer (half-integer) conformal dimension
in even (odd) space-time dimension.

The condition that the second term in (\ref{specrez})
vanishes gives the following
equations on $\sigma(s)$
\be
\label{sece}
\Big(
s^i(\lambda_i
 +\sum_{j\geq i} s^j\f{\p}{\p s^j})
+2 + \sum_{j< i}s^j(1+s^j \f{\p}{\p s^j}) \Big ) \sigma(s) \sim 0\,, \ee
where
\be
\lambda_i = l_i+\f{(d+q)}{2}+h-2i  +1\,.
\ee
Here $\sim$ implies equivalence up to
terms that are analytic in some of the variables $s^i$ and hence
do not contribute under the integral (\ref{R}).

To solve  Eq.~(\ref{sece}),
one observes that $\sigma(s)$ can be represented in the form
 \be
 \label{sigma1}
\sigma(s) = \tilde{\sigma}(s^1)\prod_{i=1}^{h}
\mu_i\left ( \frac{s^i}{s^{i+1}}\right )\,,
 \ee
where $\tilde{\sigma}(s^1)$ and $\mu_i (\frac{s^i}{s^{i+1}})$
verify the equations
\be
( s(\lambda_1 +s\frac{\p}{\p s}) +2 ) \tilde{\sigma}(s)= const\sim 0\,,
\ee
\be
\left( s^{i+1} \lambda_{i+1} +s^i (1-\lambda_i) +(s^i - s^{i+1} )
s^i\f{\p}{\p s^i} \right ) \mu_i \Big ( \frac{s^i}{s^{i+1}}\Big ) =0\,,
\ee
which is easy to solve to obtain
\be
\label{tisi}
\tilde{\sigma}(s) = \sum_{n=1}^\infty (-2)^n (\lambda_1 -n-1)! s^{-n}\q
\mu_i (x)=\sum_{n=-\infty}^\infty \frac{(\lambda_{i+1} -n -1)!}
{(\lambda_{i} -n -1)!} x^{n}\,.
\ee
Note that the issue of convergency of this infinite sum is
irrelevant because only a finite number of terms contribute to
the Lagrangian. From (\ref{tisi}) and (\ref{sigma1}) we obtain
remarkably simple explicit formula
\be
\label{sigmapow}
\sigma(s)=
\sum_{m_l\geq 0}^\infty
(-2)^{\sum_{k=1}^h m_k }
\prod_{i=1}^{h} \frac{(l_{i}+\f{(d+q)}{2} -i-1-\sum_{j=i}^h m_j)!}
{(l_{i}+\f{(d+q)}{2} -i-\sum_{j=i+1}^h m_j)!}(s^1)^{-{m_1-1}}\ldots
(s^h)^{-{m_h -1}}\,.
\ee

\subsection{Lagrangian conformal systems}
\label{lcs}
\subsubsection{General case}
Plugging Eq.~(\ref{sigmapow}) into (\ref{sigmast}) and
(\ref{R}) gives  manifest form of the  Lagrangian (\ref{lagr})
for the primary conformal fields $\Psi_{1,2}$ valued in a Lorentz tensor module
$V^{\bf l}$, that have conformal dimension $\Delta$ related to $q$ via
(\ref{confc})
\bee
\label{clag}
L^{V^{\bf l},q}(\Psi_1 , \Psi_2)&=&\sum_{m_l \geq 0}  \f{ 2^{\sum_{k=1}^h m_k }}{(q/2 - \sum_{k=1}^h m_k)!}
\prod_{i=1}^{h} \frac{l_i !(l_{i}+\f{(d+q)}{2}  -i-1-\sum_{j=i}^h m_j)!}
{ m_i ! (l_i-m_i) ! (l_{i}+\f{(d+q)}{2}  -i-\sum_{j=i+1}^h m_j)!}\nn\\
&&\ls\ls\p_{a^1_1}\ldots \p_{a^1_{m_1}} \p_{a^2_1}\ldots \p_{a^2_{m_2}}\ldots
\Psi_1^{a^1_1\ldots a^1_{m_1}c^1_{m_1+1}\ldots c^1_{l_1}\,,
a^2_1\ldots a^2_{m_2}c^2_{m_2+1}\ldots c^2_{l_2}\,,\ldots}\nn\\
&&\ls\ls\square^{q/2 - \sum_i m_i}
\p^{b^1_1}\ldots \p^{b^1_{m_1}} \p^{b^2_1}\ldots \p^{b^2_{m_2}}\ldots
\Psi_{2\,b^1_1\ldots b^1_{m_1}}{}_{c^1_{m_1+1}\ldots c^1_{l_1}\,,}
{}_{b^2_1\ldots b^2_{m_2}}{}_{c^2_{m_2+1}\ldots c^2_{l_2}
\,,\ldots}\,.
\eee
By construction, $L^{V^{\bf l},q}(\Psi_1 , \Psi_2)$ (\ref{clag}) is conformal.
Let us discuss some of its properties.

The factorials
$$\f{1}{ m_i ! (l_i-m_i) ! (q/2 - \sum_{k=1}^h m_k)!}$$
restrict summation to a finite number of terms in the region
$
0\leq m_1 \leq l_i$,    $\sum_{k=1}^h m_k\leq q/2\,,
$
where
the rest of factorials are finite for allowed Young tableaux
and $i\leq h\leq d/2$.

It should be stressed that not all terms in (\ref{clag}) are independent
because the contraction of indices of $\Psi_{1,2}$ with those of derivatives
implies their symmetrization. In particular, all differently
looking contractions of derivatives within a particular horizontal
rectangular block are equivalent up to a factor to the contraction
with its bottom row. Another property is that all terms that contain more
than $l_1$ derivatives contracted with indices of $\Psi_1$ (and hence
$\Psi_2$) are identically zero. We are not aware of  a compact
analogue of the formula (\ref{clag}) that involves only linearly
independent terms.

For
$\Psi_{1,2}=\phi^{dyn}$, where $\phi^{dyn}$ is an independent
Lorentz tensor field of conformal dimension (\ref{confc})
for a given $q\geq 0$,
conformal field  equations that follow from the Lagrangian (\ref{clag}) written
in the form (\ref{LR}) with the differential operator $E^{V^{\bf l},q}$ are
\be
\label{fl}
E^{V^{\bf l},q}\Psi_{1,2} =0\,.
\ee
In particular, in the case of trivial Lorentz representation
this gives  free Klein-Gordon equation for a scalar field
of canonical conformal dimension $\Delta = \half d-1$ as well as
the equations that contain $k^{th}$  power of the D'Alambertian
for a scalar field of conformal dimension $\Delta = \half d-k$.
 In the case  $q=0$, the operator
$E$ is a constant and the field equation (\ref{fl}) implies $\Psi_{1,2}=0$.
Numerous examples of non-gauge invariant Lagrangian conformal field equations,
that follow from the Lagrangian (\ref{clag}), were
considered in \cite{STV}.

Comparison of the obtained results with
 \cite{STV} shows that Eq.(\ref{clag}) provides  the full list
of conformal Lagrangians for conformal fields.

\subsubsection{Conformal gauge fields}

A particularly interesting case is where $\Psi_{1,2}=\phi^{dyn}$ for one of the
gauge conformal fields, \ie $\phi^{dyn}$ is valued in
the tensor space $G^{\bL,p}_{o(d-1,1)}$ and has
the conformal dimension (\ref{W1eq}).
Since the conformal dimension (\ref{W1eq}) is integer, from
(\ref{confc}) it follows that gauge invariant conformal systems only exist in even
space-time dimension. (Formally, in the case of odd $d$, the
Lagrangian becomes nonlocal containing a square root of the
second-order differential operator, that, however, does not make
much sense from various perspectives.)
{}From the results of \cite{STV} it follows that for $q>0$ there is a unique system of
differential equations on $\phi^{dyn}$ that can be Lagrangian, having as many
field equations as dynamical fields. Hence, it should coincide with the equation
(\ref{fl}) which therefore has to be gauge invariant. By (\ref{gtrp}) this means
that
\be
\label{elp}
E^{\bL}_p \cL^\bL_p=0\q q>0\,,
\ee
where $E^{\bL}_p$ is the differential operator $E$ for the dynamical field of interest.
Note that in the case of $q=0$, the equation (\ref{fl}) implies $\phi^{dyn}=0$
which is not a gauge invariant condition.

In this setup, the gauge invariance of the Lagrangian (\ref{clag})
for gauge fields $\phi^{dyn}$ is not manifest. Note that lagrangians of this
type were considered recently in \cite{mcon} for the case of symmetric conformal
fields in any dimension. To make gauge invariance manifest, one should express
the action in terms of manifestly gauge invariant objects, \ie Weyl tensors $C(\phi^{dyn})$.
Since they are also primary fields associated to the lowest weight of the
infinite dimensional Weyl module  one can try the Lagrangian (\ref{lagr})
with $\Psi_{1,2}=C(\phi^{dyn})$, \ie
\be
\label{actdy}
L^{G^{\bL,p+1}_{o(d-1,1)},q}=\half \langle C(\phi^{dyn})|{E^\bL_{p+1}}|C(\phi^{dyn})\rangle\,.
\ee
We consider the case where the dynamical field $\phi^{dyn}$
results from a $p$-form gauge field with
\be
\label{peq}
p\leq d/2 -1\,.
\ee
Since the Weyl tensor has conformal dimension (\ref{wdim}),
from (\ref{confc}) we see that $E^\bL_{p+1}$ is
a differential operator of order
\be
\label{ord}
q=d + 2L_{p+2}-2p-2\,
\ee
($d$ must be even). However, for $q>0$ the Lagrangian (\ref{actdy}) turns out
to be zero. This fact is not immediately seen from the
formula (\ref{clag}), being a consequence of the general property
of gauge invariance of its field equations of (\ref{elp}) along with (\ref{Cphi}).
Its explicit verification is
annoying even for differential forms in the simplest representations
of the conformal algebra (including the trivial representation), that we
performed  as a consistency check.

The only case where the Lagrangian (\ref{actdy})
gives rise to nontrivial field equations is that with $q=0$ which condition is
satisfied for
\be
p=d/2-1\q L_{d/2+1}=0\,.
\ee
The corresponding Lagrangian
\be
\label{actone}
L^{G^{\bL,p+1}_{o(d-1,1)},0}
=\half \langle C(\phi^{dyn})|C(\phi^{dyn})\rangle\,
\ee
gives rise to the field equations
\be \label{confeq}
 {}^*{\cL}^{\bL}_{d/2}{\cL}^{\bL}_{d/2}
 \phi^{dyn}=0\,,
\ee
where the differential operator
${}^*{\cL}^{\bL}_{p+1}$ is dual to ${\cL}^{\bL}_{p+1}$ with
respect to the integrated form (\ref{lcont}), \ie
\be \int
d^dx({}^*{\cL}^{\bL}_{p+1} A , B)= \int d^dx ( A ,{\cL}^{\bL}_{p+1}
B)\q \forall\,\, A\in G^{\bL,p+1}_{o(d-1,1)}\,,\quad B\in
G^{\bL,p}_{o(d-1,1)}\,.
\ee
 Examples
of such Lagrangians are provided by the $4d$ Fradkin-Tseytlin system \cite{FT},
where $Y$ is a two-row
rectangular Young diagram of length $s$, and its recent
generalization to any even $d$ \cite{Marnelius:2009uw}
 that corresponds in our terms
to the case of a $(d/2-1)$-form $W$ valued in a rectangular $o(d,2)$
Young diagram $Y$ of height $d/2$ and length $s$.
These are examples (\ref{sg}) and (\ref{sw}) with $h=d/2$.

Note that the Lagrangian (\ref{clag}) turns out to be dynamically nontrivial
either for the primary fields $\phi^{dyn}$ or for the Weyl tensors $C(\phi^{dyn})$,
depending on whether $E$ is a nontrivial differential operator or a constant,
respectively.

The reason why the naive application of the construction of Lagrangian
(\ref{clag}) to the Weyl tensors fails to give the correct result in the general case
is that the analysis of conformal invariance
of Section \ref{sct} does not account that gauge invariant
generalized Weyl tensors $C(\phi^{dyn})$ satisfy Bianchi identities, \ie
that the conformal invariance conditions should be imposed up to terms that are
zero by virtue of  (\ref{BI}). For the case of
a constant differential operator $E$ such terms are irrelevant because the variation
(\ref{confvar}) is trivially zero. Hence in this case the action (\ref{actone}) works properly.
For nonconstant $E$ the gauge  invariant conformal action may and,
as we have shown, should differ from the naive action (\ref{actdy}) that is trivial
in this case.

It turns out, however, that the manifestly gauge invariant conformal action for
general conformal gauge fields can be constructed in a slightly different fashion
using again the Lagrangian (\ref{clag}). The idea is to construct the action in terms
of the gauge invariant curvatures (\ref{R1}) rather than Weyl tensors.
As shown in Section \ref{stue}, from the $\gs_-$ cohomology analysis it follows
that constraints on auxiliary fields can be imposed in such a way that
the components of the gauge invariant curvatures with sufficiently low conformal
dimension vanish according to (\ref{R-}) while the lowest nonzero components
of the gauge invariant curvatures $R_C$ given by (\ref{Req}) take values
in the space $U^{\bL, p}_{o(d-1,1)}$ (\ref{U}). Once the constraints (\ref{R-}) are imposed,
$R_C$ behaves as the primary field with respect to the conformal group action
on the fiber indices. Indeed, the variation of $R_C$ under the local special
conformal transformations is proportional to the lower components $R_-$ of the
gauge invariant curvature, which are zero by (\ref{R-}). Translations, Lorentz
transformations and dilatations act on $R_L$ in the standard fashion with
$
\Delta (R_L) = -L_{p+2}\,.
$

The manifestly gauge invariant Lagrangian for any gauge conformal field associated to
the $p$-form gauge field in the representation $\bL$ of the conformal algebra
is
\be
\label{lginv}
L^{\bL,p}= g^{\un_1 \um_1} g^{\un_2\um_2}\ldots g^{\un_{p+1}\um_{p+1}}
L^{U^{\bL, p}_{o(d-1,1)},q}(R_{L\,\un_1\ldots\un_{p+1}},R_{L\,\um_1\ldots\um_{p+1}})\,,
\ee
where $\um_i$ and $\un_j$ are differential form indices contracted by the
background metric. $R_C$ is valued in the Lorentz module
$U^{\bL, p}_{o(d-1,1)}$. The differential operator $E$ still has order
$q$ (\ref{ord}).
It is important to note that Bianchi identities
do not affect the analysis of  conformal properties
of the Lagrangian (\ref{lginv}) because they account the antisymmetrization
with the differential form (\ie underlined) indices of gauge invariant curvatures
while the operator $E$ in the Lagrangian (\ref{lginv})
acts only on the fiber indices which are implicit in Eq.~(\ref{lginv}).

Plugging  the expression (\ref{CEQ})
for $R_C$ in terms of the Weyl field strength into (\ref{lginv}) we express the Lagrangian
(\ref{lginv}) in terms of $C(\phi^{dyn})$. Generically, the resulting Lagrangian
differs from (\ref{actdy}). As anticipated, the exceptional case, where the
two constructions coincide, is  $p=d/2-1$ where $E=const$.

Let us note that the form of the differential operator, that acts on the
Weyl field strengths in the resulting action, admits the ambiguity
modulo terms that vanish by virtue of the Bianchi identities
(\ref{BI}). For example, in the case of rectangular diagrams with $p=h-1$
considered in Subsection \ref{rect}, Eq.~(\ref{BI}) implies that the
antisymmetrization of any $h+1$ indices of the first derivative of the Weyl
tensor is zero (in this case, the  Bianchi identities are described by the
Young diagram that has one cell in the $(h+1)^{th}$ row in addition to the
rectangular block of height $h$). It is not hard to
see that in this case, by virtue of Bianchi identities, any conformal invariant
Lagrangian can be reduced to the form
\be
\label{lbox}
L^{G^{\bL,p+1}_{o(d-1,1)},q}= \alpha  \langle C(\phi^{dyn})|
\Box^{\frac{d}{2}-h}|C(\phi^{dyn})\rangle+ \mbox{total derivatives}\,,
\ee
where $\alpha$ is some constant.

Indeed, consider for example
the case of spin one with the Weyl tensor
\be
C_{{n_1}\,,\ldots \,,n_h}= \p_{[n_1} \phi_{{n_2}\,,\ldots \,,n_h]}\,.
\ee
By virtue of the Bianchi identities
\be
\p_{[n_0} C_{{n_1}\,,\ldots \,,n_h]}=0
\ee
 the term
\be
\p_{m}C_{{k}\,,n_2\,,\ldots \,,n_h}\p^k C^{{m}\,,n_2\,,\ldots \,,n_h}
\ee
is proportional to
\be
\p_{m}C_{n_1\,,\ldots \,,n_h}\p^m C^{n_1\,,\ldots \,,n_h}\,.
\ee
For higher spins with $s>1 $ one proceeds analogously.

Note that the Lagrangian of the type (\ref{lbox}) was considered in \cite{segal} in the
particular case of symmetric conformal fields, \ie  $h=2$. It would be interesting to see what kind of simplification
can be achieved by virtue of Bianchi identities for generic
mixed symmetry gauge conformal fields.

\subsubsection{$BF$-type conformal systems}

One can also consider the mixed case of the Lagrangian
\be
\label{actmix}
L=\half \langle \phi_1^{dyn}|{E^\bL_{p+1}}|C(\phi_2^{dyn})\rangle\,.
\ee
Again,  $L$ turns out to be trivial for $q>0$. In the case of $q=0$
this Lagrangian describes a conformal system
analogous to the $BF$ system with the Lagrangian
$L=\tilde{B}^{ab}\p_a A_b$, where $\tilde{B}^{ab} =\epsilon^{abcd}B_{cd}$
is an independent field. The equations of motion are
\be
C(\tilde{\phi}_1^{dyn})=0\q C(\phi_2^{dyn})=0\,.
\ee
However, in this setup, the gauge
symmetry of  $\tilde{\phi}^{dyn}_1$ is implicit. To describe
manifestly gauge invariant $BF$ systems it is more convenient to
use the frame-like formalism.

Consider the following $BF$-type topological action
\be
\label{BF} S=\int_{M^d} \Lambda_\Omega
(x) \wedge R_1^\Omega(W_1(x))\,,
\ee
where $R_1^\Omega$ is the gauge
invariant $(p+1)$-form curvature (\ref{R1}) built from the $p$-form
gauge field $W^\Omega_1(x)$ valued in some $o(d,2)$-tensor module
$\bL$ and $\Lambda_\Omega (x)$ is a $d-p-1$--form gauge field valued in
the same tensor module. Let $\Upsilon_\Omega$ be its $d-p$ form
curvature and $\Sigma_\Omega$ be its $d-p-2$ gauge symmetry
parameter
\be
\label{ups} \Upsilon_\Omega (x) = \D_0 \Lambda_\Omega
(x)\q \delta \Lambda_\Omega (x)= \D_0 \Sigma_\Omega (x)\,.
\ee
 This action is obviously gauge invariant under both (\ref{gtr}) and
(\ref{ups}). In addition, it is manifestly invariant under local $o(d,2)$
transformations that also act on the vacuum connection $W_0$ which
enters the covariant derivative $\D_0$. Hence it is manifestly
invariant under the global $o(d,2)$ transformations described by the
local transformations with the parameters $\epsilon_0(x)$  that
satisfy (\ref{d0}) to leave $W_0$ invariant. Note that the action
(\ref{BF}) is a simplified version of the nonlinear
action introduced in \cite{Ann} for the general (nonlinear) unfolded
system. As explained in Section \ref{weylmod}, a
 slight modification of this construction leads to the first-order
formulation of the conformal actions provided that the structure
of the full unfolded system (\ref{Weq}) and (\ref{runf}) is available.

\section{Weyl module}
\label{weylmod}

The fact that $g$ invariant field equations result
from homomorphisms of $g$--modules is well known
 (see e.g. \cite{Kos,ER,Bast_East} and references
therein). The novelty of the unfolded
dynamics approach is that it incorporates  gauge
fields and gauge symmetries via inclusion of $p>0$ differential
forms. This extension is of primary
importance for the analysis of physically interesting
gauge invariant nonlinear models.

\newcommand{\subplus}
{\begin{picture}(12 ,6 )(3,0) \put(0,-.5){ $\subset$ }
\put(7,02.5){\line(1,0){5}} \put(9.5,0){\line(0,1){5}}
\end{picture}}

In this section we comment briefly on the relation of the results of
this paper with those of \cite{STV}, where the classification
of conformal invariant field equations in any dimension was given
based on the analysis of generalized Verma modules. Recall that
generalized Verma modules are induced from finite dimensional
modules of the parabolic subalgebra $p=h\subplus t$ that extends the
maximal compact subalgebra $h=o(2)\oplus o(d)\subset o(d,2)$ by the
subalgebra of translations $t$ spanned by $P_a$. A generalized Verma
module $V_h$ is therefore characterized by the weights of $h$. The
latter characterize a tensor type and conformal dimension of the
dynamical field $\phi^{dyn}(x)$.

The unfolded equations studied in \cite{STV} have the form
\be
\label{uneq}
\D\Phi(x)=0\,,
\ee
where $\Phi(x)$ is a zero-form field valued in
$V_h$ and $\D$ is the $o(d,2)$ covariant derivative in $V_h$.
$\phi^{dyn}(x)$ are components of $\Phi(x)$ valued
in the vacuum subspace  from which $V_h$ is induced.

The two cases are different. If $V_h$ is irreducible,
the equation (\ref{uneq}) is a set of constraints that
express all components of $\Phi(x)$ via derivatives of
$\phi^{dyn}(x)$. If $V_h$ is reducible, it contains a submodule
that is also a generalized Verma module $V_{h^\prime}$
induced from a singular space $S\in V_h$ that itself forms
a $p$-module being annihilated by the translation generators.
The appearance of a submodule (singular subspace) in
$V_h$ implies that the equation (\ref{uneq}) contains
a differential equation on $\phi^{dyn}(x)$ and  another
dynamical field $\phi^{\prime dyn}(x)$ valued in $ S$ appears,
that is not expressed via derivatives of $\phi^{dyn}(x)$
(and hence the equation on $\phi^{dyn}(x)$).

If unfolded equations are imposed
on fields valued in some $g$-module $V$ that
contains a submodule $U$, \ie the sequence
\be
0\rightarrow U\rightarrow V\rightarrow W\rightarrow 0\,
\ee
is exact, then a set of equations on the fields $\phi$
valued in the quotient module
$W=V/U$ constitutes  a subsystem of the unfolded equations while the
set of equations on the fields $\chi$ valued in  $U$ may
receive a contribution from the fields valued in $W$, \ie
schematically,
\be
\label{phi}
d\phi +W_0 \phi =0\,,
\ee
\be
\label{chi}
d\chi + W_0 \chi +W_0 \phi =0.
\ee
For a reducible generalized Verma module $V$,
the equation (\ref{phi}) contains a nontrivial differential
condition.

To go off-shell it is more
convenient to have opposite structure with the roles
of $\phi$ and $\chi$ exchanged. This is achieved in the contragredient
module  $\CV$ which is the dual module to $V$ with the exchanged roles
of translations and special conformal transformations (\ie the dualization
is combined with Chevalley  involution). The structure of $\CV$ is
dual to that of $V$, \ie
\be
0\rightarrow W^\natural\rightarrow V^\natural
\rightarrow U^\natural\rightarrow 0\,,
\ee
where $W^\natural$ and $U^\natural$ are dual to $W$ and $U$
respectively. Correspondingly, the equations dual to (\ref{phi})
and (\ref{chi}) have the structure
\be
\label{phinat}
d\phi^\natural +W_0 \phi^\natural +W_0 \chi^\natural  =0\,,
\ee
\be
\label{chinat}
d\chi^\natural + W_0 \chi^\natural  =0.
\ee
The difference between the systems (\ref{phi}),(\ref{chi})
and (\ref{phinat}),(\ref{chinat}) is that the appearance of
$\chi^\natural$ in (\ref{phinat}) puts this equation off-shell,
expressing $\chi^\natural$ via left hand sides of the field
equations resulting from Eq.~(\ref{phi}).
Provided that $\chi^\natural$ is valued in an irreducible
module $U^\natural$, the equation (\ref{chinat})
imposes no differential conditions on $\chi^\natural$ apart from
Bianchi conditions that may follow from (\ref{phinat}).

In the non-gauge case the
structure of $\CV$ is just as described. The fields $\chi^\natural$
describe the left hand sides of the
field equations on $\phi^\natural$ and their derivatives.
Setting $\chi^\natural=0$,
one puts the system on-shell. The resulting irreducible
$o(d,2)$-module $W^\natural$ where $\phi^\natural$ is valued
consists of various on-shell nontrivial derivatives of the
dynamical field $\phi^{\natural dyn}(x)$.
The case of series of generalized Verma modules listed in Eqs.
(4.46), (4.66) of \cite{STV} corresponds to the nongauge conformal systems.

On the other hand, the pattern of Lorentz representations
and conformal weights of the  zero-forms valued in $H^p(\gs_-,
F^\bL_{o(d,2)})$ matches the list of weights of generalized
Verma modules (4.41), (4.49) of \cite{STV}. These are the series of
generalized Verma modules that are based on dominant integral
highest weights of
$o(d,2)$.
The latter property agrees with the analysis of this paper performed
in terms of gauge fields valued in finite dimensional
$o(d,2)$--modules, \ie dominant integral highest weights.
In this case the structure of $\CV$ is more subtle.

Let $V$ be induced from the module equivalent to that of the Weyl tensor
associated to some  $o(d,2)$ weight $\bL$ and
differential form degree $p$
(being implicit, $\bL$
and $p$ are assumed to be fixed in this section). By definition,
$\CV$  is fully off-shell in the sense that  the covariant constancy
equation
\be D \cC^\ga (x) = 0
\ee
on the fields $\cC^\ga (x)$ valued in $\CV$
imposes no restrictions on its ground state field identified with
the generalized Weyl tensor $C(x)$, merely expressing all components of
$\cC^\ga (x)$ via derivatives of  $C(x)$. This
pattern, however does not take into account that $C(x)$ should obey the
Bianchi identities as a consequence of the equation (\ref{Req}) or,
equivalently, that $C(x)$ can be built from the dynamical fields according to
(\ref{Cphi}). Demanding the Weyl tensor to satisfy Bianchi
identities implies that some combinations of derivatives of the Weyl
tensor are set to zero. Since the setting is conformal invariant,
this set of derivatives should be determined by the unfolded field
equations for the components valued in some quotient module
$\B=\CV/W^{off}$ where $W^{off}$ is a submodule of $\CV$.
Roughly speaking, the module $\CV$ is spanned by
various derivatives of the unrestricted Weyl tensor $
 \cR C\,,$ where $\cR$ is an arbitrary  differential operator,
while the fields valued in the off-shell  Weyl submodule $W^{off}$
describe those derivatives of the Weyl tensor that remain nonzero
upon $C(x)$ is expressed in terms of the dynamical field
$C(x)=\cL \phi^{dyn}(x)$.
The module $W^{off}$ is off-shell in the sense that the unfolded
equations (\ref{Weq}) and (\ref{runf}) impose no conditions on the
dynamical field $\phi^{dyn}$. Let us stress that the module $\CV$
has just one nontrivial submodule \cite{STV} which therefore is $W^{off}$.

The crucial question then is whether or not it is possible to impose
further conformal invariant differential equations on $C$. The full
answer to this question can be read off the results of \cite{STV}
and turns out to be surprisingly simple. The cases of odd and even
dimensions are essentially different in this respect.

In odd dimension $d=2k+1$, $W^{off}$ turns out to be
irreducible, which means that, except the trivial (pure gauge) case
of the equation $C(\phi(x))=0$ no conformal invariant field equations can be imposed
on the Weyl tensor for any type of conformal gauge fields. This
implies that any gauge invariant system in odd dimensions is trivial,
being either off-shell with the Weyl tensor restricted only by
Bianchi identities or pure gauge with zero Weyl tensor. For example,
requiring the Weyl tensor to vanish in conformal gravity implies
that the metric tensor is conformally flat, hence being pure gauge
with respect to local dilatations.

 In even dimensions, $W^{off}$ contains a single
submodule $W^{on}$ while $W^{off}/W^{on}$ is
 spanned by various derivatives of the left-hand sides
of the field equations. The restriction of $W^{off}$ to $W^{on}$
just imposes the nontrivial equations on the Weyl tensor $C(\phi^{dyn})$ and
hence on $\phi^{dyn}$. Note that the appearance of nontrivial field equations
in this case is due to so-called subsingular space in the generalized
Verma module. This means
that although the module $V$ has just one submodule $V^\prime$, the
quotient module $V/V^\prime$ turns out to be reducible, \ie it acquires
upon factorization a singular subspace called subsingular space. In
more physical terms this means that nontrivial equations on the Weyl tensor
can only be imposed along with Bianchi identities which hold
automatically once it is expressed in terms of derivatives of the
dynamical field $\phi^{dyn}$. In fact, this implies that the
equations are imposed on $\phi^{dyn}$. These are
the equations that follow from the Lagrangian (\ref{lginv}).

Let us stress the difference between  two seemingly similar classes of
conformal HS fields.

One class consists of unitary conformal modules associated to
$V_h$ where $h$ corresponds to rectangular Young tableaux
of height $d/2$ (hence $d$ is even), length $s$ and conformal
dimension $d/2-s+1$. These systems were considered in
\cite{town,Siegel,BBAST,gr} and correspond to the
fields in the middle of the rhomb in the diagram
in ({B.2}) of \cite{STV} with $\lambda_i=s-1$, $q=d/2$.

Another class consists
of the conformal gauge fields, that correspond to $(d/2-1)-$forms
valued in rectangular Young diagrams of height $d/2$
\cite{FT,Marnelius:2009uw}.
In notation of \cite{STV} their Weyl tensors  are described
by the weights $(-d/2, s,s,s\ldots \pm s )$. Since the first weight
is minus conformal dimension \cite{STV}, we
observe that the two types of fields
have different conformal dimensions for all spins
 $s\neq 1$. For $s=1$  they coincide describing the case of
forms.

 For all $s>1$, the two types of systems are
essentially different. The gauge system is
non-unitary while the unitary system is not gauge, which means
that the representation of the fields to be associated to HS Weyl tensors
as derivatives of gauge potentials in the unitary case breaks conformal
invariance because the conformal dimension does not match.
(Note, however, that in \cite{33} it was shown
that there is a nonstandard possibility to extend
the unitary system to the gauge case by such a doubling of the system
in Anti-de Sitter background that is singular in the flat limit.)

The actions obtained in Section \ref{act} have a form typical for the
metric-like formalism, although the structure of both
dynamical fields and gauge invariant field strengths was derived
from the unfolded dynamics based on the frame-like formalism that
operates with differential forms. Since the
frame-like formalism seems to be most appropriate for the extension
to the nonlinear case, it is interesting whether there is a
natural formulation for conformal actions in the frame-like formalism.
A special feature of conformal gauge theories
is that the nontrivial cohomology responsible for
field equations belongs to the Weyl module rather than to the gauge
field module as in the unitary HS models. As a result, the corresponding
actions should be built in terms of  the
Weyl module. In this respect they should be analogous to
the scalar field action proposed in \cite{sigma}.
More precisely, we conjecture that the corresponding action has the structure
\be \label{first} S=\int_{M^d} \Big (\Lambda
(x) \wedge (R_1(W_1(x))-
\underbrace{W_0^{\cdots}\wedge\ldots W_0^{\cdots}}_{p+1} \cC)
+\lambda \D\cC + \half \langle C|{E}|C\rangle\Big )\,,
\ee
where $C$ is regarded as an independent field and $\cC$ denotes
the infinite set of fields valued in $W^{off}$.
 The equations for the Lagrange
multipliers impose the off-shell unfolded equations that
express the Weyl tensor $C$ via derivatives of the dynamical fields
$\phi^{dyn}$ and $\cC$ via derivatives of $C$.
Substitution of this result into the last term reproduces the action (\ref{lginv}).

The action (\ref{first}) is expected  to provide a starting point
towards a nonlinear HS conformal action within the
unfolded dynamics approach. For the case of symmetric HS
fields, it would be interesting to compare the
result with the nonlinear action proposed by Segal \cite{segal} and
 with the free second-order action proposed by Metsaev
\cite{Metsaev:2007rw,Metsaev:2007fq}. To this end, it remains however
to derive the manifest form of
the full unfolded system (\ref{Weq}) and (\ref{runf}).

\section{Conclusion}
\label{conclusion}

In this paper  conformal mixed symmetry gauge fields
in any dimension were analyzed in the frame-like approach.
In particular, the dynamical content of the gauge system based on
$p$-form gauge fields in any finite dimensional tensor module of the
conformal algebra $o(d,2)$  was
worked out. This allowed us to describe a variety of
conformal systems that include all known and infinitely
many new mixed symmetry conformal gauge fields. Conformal
invariant actions are constructed for generic conformal
systems that include both gauge and  non-gauge fields.
Comparison of the obtained results with the list of nontrivial
conformal field equations of \cite{STV} suggests that
the list of bosonic conformal Lagrangian systems presented in
this paper is complete.

Technically,  the analysis of the
dynamical content of a system is performed in terms
of so-called $\gs_-$ cohomology \cite{sigma} which is
particulary simple in  the conformal case where it amounts to
the Lie algebra cohomology.
Although the latter can be read of the
literature (see, e.g., \cite{STV} and references therein),  its
analysis is presented in full generality because the applied method,
based on the analysis of supersymmetric vacua in a
specific supersymmetric matrix model, is quite efficient and can
have other applications.
Every supersymmetric state gives rise to a nontrivial $\gs_-$
cohomology that has clear dynamical interpretation in conformal
field-theoretical models in terms of dynamical gauge fields, gauge
invariant field strengths, gauge symmetry parameters, Bianchi
identities, syzygies and gauge symmetries for gauge symmetries.

The obtained results are anticipated to have  applications
to the description of more complicated unitary field-theoretical
systems in $AdS_d$ space, which are of most physical interest. A
useful viewpoint is to interpret nonconformal systems in $AdS_d$ as
conformal models with the $o(d,2)$ symmetry broken to the $AdS_d$
symmetry $o(d-1,2)$ somewhat in spirit of two-time physics \cite{B}.
 This approach looks
promising for better understanding of the dynamical content of
$AdS_d$ dynamical systems. In this case, the $o(d,2)$ irreducible
multiplets of fields of conformal theory decompose into reducible
sets of $o(d-1,2)$ multiplets, some of which may be set to zero by
the conditions that break down the $o(d,2)$ symmetry. One
implication of this phenomenon is that the $AdS$ $\gs_-$ cohomology
should result from two sources. One comes from the conformal
theory cohomology, \ie Weyl field strengths.
Indeed, the Weyl tensors of the conformal models are shown to
coincide with those of the $AdS$ mixed symmetry systems
\cite{ASV1,ASV2}. Another part results from the comparison
of the sets of  gauge fields in the $AdS$ models and those in
the conformal models.
Along these  lines, it should be possible to give a concise
interpretation of recent computation by Skvortsov \cite{skads,skads1} of
$\gs_-$ cohomology in the $AdS$ models of mixed symmetry unitary
fields.

Other directions for the future research include the analysis of
fermionic conformal systems as well as of generalized conformal
systems based on $sl_n$ (see also  \cite{skads1}) and $sp(2M)$. The
$sl_n$ models  are expected to be related to off-shell
field-theoretical systems (see e.g. \cite{BBAST} and references therein),
while the $sp(2M)$ models are of interest for the  models in the
$sp(2M)$ invariant matrix space-times
\cite{F,BLS,BHS,Mar,GV,33} (and references therein).

Finally, let us stress that unfolded dynamics provides a particularly
useful tool for uplifting the field-theoretical models to larger spaces
where their symmetries are geometrically realized.
For conformal models considered in this paper, an interesting
development would be their reformulation in $d+2$ dimensional
space-times where the conformal group $O(d,2)$ acts geometrically,
extending the previous developments along this direction  \cite{dcon,Marn,con,B}.

\section*{Acknowledgments}

I am grateful to O.Gelfond and E.Skvortsov for helpful comments on the
manuscript and to M.Grigoriev,  D.Sorokin and, especially,
O.Shaynkman for useful discussions.
This research was supported in part by RFBR Grant No 08-02-00963,
LSS No 1615.2008.2 and Alexander von Humboldt Foundation Grant PHYS0167.

\section*{Appendix\\ Properties of the conformal invariance condition}
Using (\ref{comlor}), the equation
\be \label{specg} E^b M_{ba} +\half E_b{}^b \p_a
=\gamma E_a\,,
\ee
where $\gamma$ is a free parameter, can be equivalently
written in the form
\be \label{spec1} M_{ab} E^b +\half E_b{}^b \p_a   =\tilde{\gamma}
E_a\,,
\ee
where
\be \label{tilga} \tilde{\gamma}=d+q-2-\gamma\,
\ee
and $q$ is a degree (\ref{degr}) of the differential operator $E$.
Let $E_1$ and $E_2$ be solutions of (\ref{specg}) with some
parameters $\gamma_1$ and $\gamma_2$. It is easy to see that
\be (E_1 E_2)^b M_{ba} +\half (E_1 E_2)_b{}^b
\p_a = (\gamma_2 +q_1)E_1 E_{2a} +\gamma_1 E_{1a} E_2\,,
\ee
where
$q_1$ is a degree of the differential operator $E_1$.
As a result, the operator $E_{1,2}=E_1 E_2$ satisfies the equation
(\ref{specg}) with $\gamma=\gamma_1$ provided that
\be
\gamma_1=\gamma_2 +q_1\,. \ee
By virtue of
(\ref{tilga}), this is equivalent to
\be
\tilde{\gamma}_2=\tilde{\gamma}_1 +q_2\,.
\ee

\end{document}